\documentclass[twocolumn,floats,amsmath,amssymb,pre]{revtex4}

\usepackage{graphicx}

\begin{document}

\title{Ising model in clustered scale-free networks}
\author{Carlos P. Herrero}
\affiliation{Instituto de Ciencia de Materiales,
         Consejo Superior de Investigaciones Cient\'ificas (CSIC),
         Campus de Cantoblanco, 28049 Madrid, Spain} 
\date{\today}

\begin{abstract}
The Ising model in clustered scale-free networks has been studied by 
Monte Carlo simulations. These networks are characterized by a degree
distribution of the form $P(k) \sim k^{-\gamma}$ for large $k$. 
Clustering is introduced in the networks by inserting triangles,
i.e., triads of connected nodes.
The transition from a ferromagnetic (FM) to a paramagnetic (PM) phase
has been studied as a function of the exponent $\gamma$ and the
triangle density.
For $\gamma > 3$ our results are in line with earlier simulations, and
a phase transition appears at a temperature $T_c(\gamma)$ in the 
thermodynamic limit (system size $N \to \infty$).
For $\gamma \leq 3$, a FM--PM crossover appears at a
size-dependent temperature $T_{\rm co}$, so that the system remains in 
a FM state at any finite temperature in the limit $N \to \infty$.
Thus, for $\gamma = 3$, $T_{\rm co}$ scales as $\ln N$, whereas
for $\gamma < 3$, we find $T_{\rm co} \sim J N^z$, where the 
exponent $z$ decreases for increasing $\gamma$. Adding motifs 
(triangles in our case) to the networks causes an increase in the
transition (or crossover) temperature for exponent $\gamma > 3$ (or $\leq 3$).
For $\gamma > 3$, this increase is due to changes in the mean values
$\langle k \rangle$ and $\langle k^2 \rangle$, i.e., the transition is
controlled by the degree distribution (nearest neighbor connectivities).  
For $\gamma \leq 3$, however, we find that clustered and unclustered
networks with the same size and distribution $P(k)$ have different 
crossover temperature, i.e., clustering favors FM correlations,
thus increasing the temperature $T_{\rm co}$.
The effect of a degree cutoff $k_{\rm cut}$ on the asymptotic behavior
of $T_{\rm co}$ is discussed.  
\end{abstract}



\maketitle

\section{Introduction}

Many natural and artificial systems have a network structure,
with nodes representing typical system units and edges playing the role of
interactions between connected pairs of units.
Complex networks can be used to model various kinds of real-life
systems (social, economic, technological, biological), and to analyze 
processes taking place on them \cite{st01,al02,do03,ne10,co10}.
In recent years, various network models have been designed capturing 
aspects of real systems, thus allowing to explain empirical data 
in several fields.
This is the case of small-world \cite{wa98} and scale-free 
networks \cite{ba99b}, which provide us with the underlying topological 
structure to analyze processes such as signal propagation \cite{wa98,he02b},
as well as the spread of information \cite{pa01,la01b}, opinions \cite{ca07b}
and infections \cite{mo00,ku01}. 
These types of networks have been also used to study statistical 
physical problems as percolation \cite{mo00,ne99} and 
cooperative phenomena \cite{ba00,le02,he02,ca05,do08,he09b,os11}. 

In scale-free (SF) networks the degree distribution $P(k)$, where $k$ is the
number of links connected to a node, has a power-law decay 
$P_{\rm sf}(k) \sim k^{-\gamma}$ \cite{do02,go02}.
This type of networks have been found in several real-life systems,
such as the internet \cite{si03}, the world-wide web \cite{al99}, 
protein interaction networks \cite{je01}, and social systems \cite{ne01}.
In both natural and artificial systems, the exponent $\gamma$ controlling the
degree distribution is usually in the range $2 < \gamma < 3$ \cite{do02,go02}. 
The origin of power-law degree distributions was studied by
Barab\'asi and Albert \cite{ba99b}, who found that two ingredients
can explain the scale-free nature of networks, namely growth and 
preferential attachment. More general models based on these ingredients
have appeared later in the literature \cite{kr00,he11b}. 
One can also deal with equilibrium SF networks, defined as statistical
ensembles of random networks with a given degree distribution 
$P_{\rm sf}(k)$, for which one may analyze several properties as a function 
of the exponent $\gamma$ \cite{do02,bo06}.

Many real-life networks include clustering, i.e., the probability of
finding loops of small size is larger than in random networks.
This has been in particular quantified by the so-called clustering 
coefficient, which measures the likelihood of
``triangles'' in a network \cite{ne10}.
Most network models employed in the past did not include clustering.
Some of them, such as the Watts-Strogatz small-world model \cite{wa98} 
show clustering, but are not well-suited as models of most actual networks. 
Several computational models of clustered networks have been defined 
along the years \cite{ho02,kl02,se05}, but in general their properties 
cannot be calculated by analytical procedures.
In last years, it was shown that generalized random graphs can be
generated incorporating clustering in such a way that exact formulas can
be derived for many of their properties.
This is the case of the networks defined by Newman \cite{ne09}
and Miller \cite{mi09}.

Cooperative phenomena in complex networks are known to display 
characteristics related to the particular topology of these systems
\cite{do08}.  The Ising model on SF networks has been studied by using 
several theoretical techniques \cite{le02,do02b,ig02,he04,do10,me11b}, 
and its critical behavior was found to depend on the exponent $\gamma$. 
Two different regimes appear for uncorrelated networks. 
On one side, for an exponent $\gamma > 3$, the average value 
$\langle k^2 \rangle$ is finite in the large-size limit, and there 
appears a ferromagnetic (FM) to paramagnetic (PM) transition at a finite 
temperature $T_c$.  
On the other side, when $\langle k^2 \rangle$ diverges 
(as happens for $\gamma \leq 3$), the system remains in its ordered 
FM phase at any temperature, so that there is no phase transition in the
thermodynamic limit. 
The antiferromagnetic Ising model has been also studied in scale-free 
networks, where spin-glass phases have been found \cite{ba06,he09}.

All this refers to unclustered random networks with a power-law degree
distribution.
One may ask how this picture changes when the networks are clustered,
i.e., the clustering coefficient has a non-negligible value.
In principle, one expects that the presence of small loops in the 
networks will enhance correlations between spins located on
network nodes, thus favoring ordered schemes such as an FM pattern.
Thus, the effects of clustering on various cooperative phenomena in 
complex networks have been studied earlier. This is the case of 
percolation \cite{gl09,gl10,ka10b,al12}, epidemics \cite{mi09,wa12,mo12},
and dynamical processes \cite{he10,me11}. 
Yoon {\em et al.} \cite{yo11} studied the Ising model on networks with
arbitrary distribution of motifs using an analytical procedure, the
so-called belief-propagation algorithm. For the networks considered in 
that paper, where the thermodynamic limit is well defined 
($T_c$ converges for $N \to \infty$), these authors found 
that clustering increases the critical temperature in comparison with 
tree-like networks with the same mean degree, but does not change
the critical behavior. 

In this paper we study the FM-PM transition for the Ising model in 
scale-free networks with clustering, which is realized by introducing
triangles in the networks, i.e., three-membered loops. 
Several values of the exponent $\gamma$ are considered, as well as 
various concentrations of triangles. 
We employ Monte Carlo (MC) simulations to obtain the transition 
temperature, when it is well defined (for $\gamma > 3$ ), and  to derive
the size dependence of the crossover temperature for cases where
$\langle k^2 \rangle$ diverges as $N \to \infty$ ($\gamma \leq 3$).  

The paper is organized as follows.
In Sec.~II we describe the clustered networks considered here.
In Sec.~III we present the computational method employed to carry out 
MC simulations.
In Sec.~IV we present results of the simulations and a discussion
for the different parameter regions, depending on the value of the
exponent $\gamma$ ($> 3$, $= 3$, or $< 3$).
The paper closes with a summary in Sec.~V.

\section{Scale-free networks with clustering}

We consider clustered networks with a degree distribution
$P(k)$ that follows the power-law dependence 
$P_{\rm sf}(k) \sim k^{-\gamma}$ for large degree $k$.
Clustering is included by inserting triangles in the 
networks, i.e., triads of connected nodes.
Other kinds of polygons (squares, pentagons, ...) can be introduced
to study their effect on critical phenomena in physical systems, but 
we choose triangles since they cause stronger correlations 
between entities defined on network sites, as in the case of the Ising
model considered here. 

We generate networks by following the method proposed by Newman \cite{ne09},
where one separately specifies the number of edges and the number of 
triangles.
This procedure allows one to generalize random graphs to incorporate 
clustering in a simple way, so that exact formulas can be derived for 
many properties of the resulting networks \cite{ne09}.

\begin{figure}
\vspace{0.2cm}
\includegraphics[width=7cm]{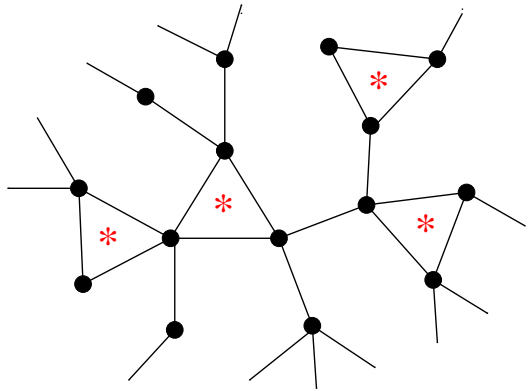}
\vspace{0.0cm}
\caption{
Schematic representation of a typical network considered in this
work, for which one separately specifies the number of single links
and triangles attached to each node. Triangles are indicated by
asterisks (*).
}
\label{f1}
\end{figure}

For a network of size $N$ (number of nodes), we call $t_i$ the number of 
triangles in which node $i$ takes part, and $s_i$ the number of single edges
not included in the triangles. This means that edges within the triangles
are listed separately from single links.
Thus, a single link can be viewed as a network element joining together
two nodes and a triangle as an element connecting three nodes.
The degree $k_i$ of node $i$ is then $k_i = s_i + 2 \, t_i$, as
each triangle connects it to two other nodes.
A picture of such a network is presented in Fig.~1, where
triangles are indicated with asterisks (*).

To generate the networks, we first define the edges.  
We assign to each node $i$ a random number 
$s_i$, which represents the number of outgoing links from this
node (stubs). The set of numbers $\{ s_i \}_{i=1}^N$ (with $s_i \geq k_0$,
the minimum allowed degree) is taken from the probability distribution 
$P_{\rm sf}(s) \sim s^{-\gamma}$ \cite{ne05}, giving
a total number of stubs $K = \sum_{i=1}^N s_i$.  
We impose the restriction that $K$ must be an even integer.
Then, we connect stubs at random
(giving a total of $L = K/2$ connections), with the conditions:
(i) no two nodes can have more than one bond connecting them
   (no multiple connections), and
(ii) no node can be connected by a link to itself (no self-connections).

In a second step we introduce triangles into the networks. 
Their number $N_{\Delta}$ is controlled by the parameter $\nu$, which gives 
the mean number of triangles in which a generic node is included 
($N_{\Delta} = \frac13 N \nu$). 
The number of triangles $t_i$ associated to a node $i$ is drawn from
a Poisson distribution $Q(t) = {\rm e}^{-\nu} \nu^t / t!$.
Thus, we have $t_i$ ``corners'' associated to node $i$, and the total
number is $T = \sum_{i=1}^N t_i = 3 N_{\Delta}$.
We impose the condition that $T$ be a multiple of 3.
Then, we take triads of corners uniformly at random to form
triangles, taking into account conditions (i) and (ii) above 
to avoid multiple and self-connections.
Note that single links can by chance form triangles. Calling $N_t$ the 
number of such triangles, their density $N_t/N$ for a given mean degree 
$\langle k \rangle$ vanishes as $N \to \infty$.
In fact, $N_t / N$ scales as $1/N$ for large $N$ \cite{ne09,ne10,yo11}.

In complex networks, one usually defines the clustering coefficient
$C$ as the ratio $C = 3 N_{\Delta} / N_3$, where $N_3$ is the number
of connected triplets \cite{ne10}:
\begin{equation}
 N_3 = N \sum_k  \frac {k (k-1)}{2} \, P(k) = 
   \frac12 N \left( \langle k^2 \rangle - \langle k \rangle \right)  \, .
\end{equation}
Thus, for the networks discussed here, we have
\begin{equation}
    C = \frac {2 \nu}{ \langle k^2 \rangle - \langle k \rangle } \; ,
\end{equation}
and the clustering coefficient can be changed as a function of the 
parameter $\nu$.

\begin{figure}
\vspace{-1.0cm}
\includegraphics[width= 8cm]{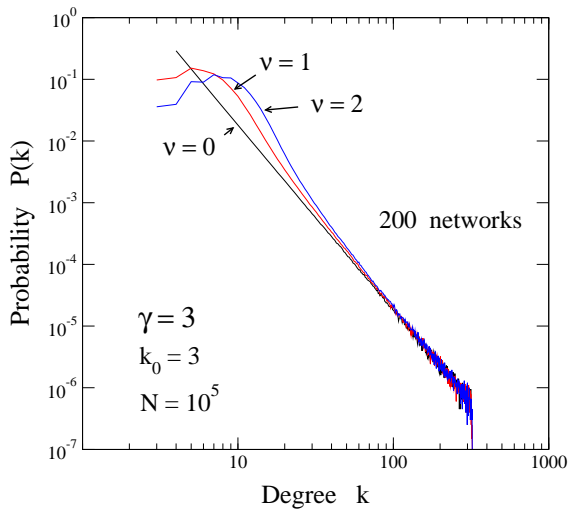}
\vspace{-0.3cm}
\caption{
Probability density as a function of the degree $k$ for networks
with $\gamma = 3$, minimum degree $k_0$ = 3, and $N = 10^5$ nodes.
The data shown are an average over 200 network realizations for each
value of the parameter $\nu$ = 0, 1, and 2.
}
\label{f2}
\end{figure}

Aside from $\gamma$ and the triangle density $\nu$, 
our networks are defined by the minimum degree $k_0$. 
Since we are interested in finite-size effects, the network
size $N$ is also an important variable in our discussion.
The degree distribution $P(k)$ obtained for networks generated by
following the procedure described above is presented in Fig.~2. 
In this figure, we have plotted $P(k)$ for networks including $10^5$
nodes, with $\gamma = 3$ and $k_0 = 3$. Each curve corresponds to a 
particular value of the parameter $\nu$ = 0, 1, and 2, including in each 
case an average over 200 network realizations.
Comparing the curves corresponding to different $\nu$ values, one observes
that the introduction of triangles in the networks causes clear changes
in the distribution $P(k)$ for small $k$ values. However, for large 
degrees, the distribution is found to follow the dependence
$P_{\rm sf}(k) \sim k^{-\gamma}$ typical of scale-free networks, with 
$\gamma = 3$ in the present case.
This could be expected, since the Poisson distribution $Q(t)$ associated
to the triangles has a much faster exponential-like decay for large $t$.
In the three cases shown in Fig.~2 there appears an effective
cutoff $k_{\rm cut} \gtrsim 300$, which is related to the finite size $N$
of the networks (see below).
We note that a maximum degree $k_{\rm cut}$ was explicitly introduced earlier 
in scale-free networks for computational convenience \cite{he04}. 

An important characteristic of the considered networks, that will be
employed below to discuss the results of the Ising model, is the
mean degree $\langle k \rangle$.
For scale-free networks with $\nu = 0$, the mean degree is given by 
\begin{equation}
 \langle k \rangle_{\infty} = \sum_{k=k_0}^{\infty} k \, P_{\rm sf}(k) 
        \approx   k_0 \,  \frac{\gamma-1}{\gamma-2}   \, ,
\label{kmeaninf}
\end{equation}
where the last expression is obtained by replacing the sum by an integral,
which is justified for large $N$.
Note that we assume here $\gamma > 2$ and that the distribution
$P_{\rm sf}(k)$ is normalized to unity (for $\gamma \leq 2$ the mean degree
$\langle k \rangle$ diverges in the large-size limit: 
$\langle k \rangle_{\infty} \to \infty$).
Then, for our networks including triangles, one has
\begin{equation}
     \langle k \rangle_{\infty}   \approx   
            2 \, \nu +  k_0 \, \frac{\gamma-1}{\gamma-2}    \, .
\label{kmean}
\end{equation}

\begin{figure}
\vspace{-1.0cm}
\includegraphics[width= 8cm]{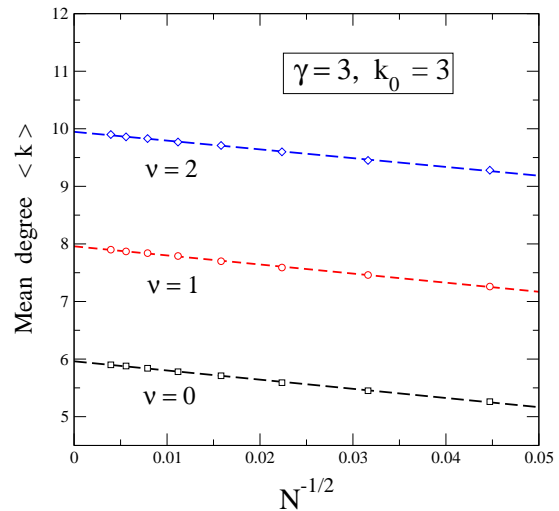}
\vspace{-0.3cm}
\caption{
Mean degree $\langle k \rangle$ as a function of $N^{-1/2}$ for
networks with $\gamma = 3$, $k_0 = 3$, and three values of
$\nu$: 0 (squares), 1 (circles), and 2 (diamonds).
Error bars of the simulations results are less than the symbol size.
Lines are fits to the expression
$\langle k \rangle = a + b \, N^{-1/2}$.
}
\label{f3}
\end{figure}

For finite networks, a size effect is expected to appear in the
mean degree, as a consequence of the effective cutoff appearing in the
degree distribution (see Fig.~2).
This is shown in Fig.~3, where we present $\langle k \rangle$ vs $N^{-1/2}$, 
for our generated networks with $\gamma = 3$ and $k_0 = 3$.
The mean degree decreases as $N^{-1/2}$ increases, i.e., $\langle k \rangle$
 increases as the system size is raised. 
The data shown in Fig.~3 for $\nu$ = 0, 1, and 2, follow a linear
dependence, and in fact can be fitted as 
$\langle k \rangle = a + b \, N^{-1/2}$, at least in the region plotted in 
the figure ($N > 500$). 
The fit parameter $a$ is close to the mean degree $\langle k \rangle_{\infty}$ 
given in Eq.~(\ref{kmean}), and the small difference is mainly due to 
the replacement of sums by integrals in the derivation of that equation.
Note that for $\gamma = 3$, $\nu = 0$ and $k_0 = 3$, Eq.~(\ref{kmean}) yields
$\langle k \rangle_{\infty}$ = 6.
The slope $b$ obtained from the linear fits turns out to be the same
(within statistical noise) for the three cases shown in Fig.~3.

This size dependence of $\langle k \rangle$ can be understood by noting 
that the effective cutoff
$k_{\rm cut}$ appearing in a power-law degree distribution is related with
the network size $N$ by the expression \cite{do02b,ig02}
\begin{equation}
   \sum_{k_{\rm cut}}^{\infty} P_{\rm sf}(k) = \frac{c}{N}  \; ,
\label{sumkc}
\end{equation} 
where $c$ is a constant on the order of unity.
From this expression, one can derive for $\gamma = 3$ (see the Appendix):
\begin{equation}
  \langle k \rangle \approx \langle k \rangle_{\infty}
       \left[ 1 - \left( \frac{c}{N} \right)^{\frac12} + 
        {\cal O} \left( \frac1N \right)     \right]  \; .
\label{kmean3}
\end{equation}
Comparing with the fit shown in Fig.~3, one finds $c \approx 7$,
which introduced into Eq.~(\ref{kcutk}) yields for the cutoff
$k_{\rm cut} \approx 360$, in line with the results shown in Fig.~2,
thus providing a consistency check for our arguments.

For scale-free networks with an exponent $\gamma < 3$, 
Catanzaro {\em et al.} \cite{ca05b} found that appreciable correlations
appear between degrees of adjacent nodes when no multiple and
self-connections are allowed. Such degree correlations can be avoided 
by assuming a cutoff $k_{\rm cut} \sim N^{1/2}$. Thus, for
$\gamma < 3$ we generate here networks with cutoff
$k_{\rm cut} = N^{1/2}$.
For the clustered networks considered here, generated as in Ref.~\cite{ne09},
the presence of triangles introduces degree correlations between nodes 
forming part of a triangle in the network.

\section{Simulation method}

On the networks defined in Sec.~II, we consider a spin model given
by the Hamiltonian
\begin{equation}
H = - \sum_{i < j} J_{ij} S_i S_j   \, ,
\end{equation}
where $S_i = \pm 1$ ($i = 1, ..., N$) are Ising spin variables, 
and the coupling matrix $J_{ij}$ is given by
\begin{equation}
\label{Jij}
J_{ij}  \equiv \left\{
     \begin{array}{ll}
         J (> 0), & \mbox{if $i$ and $j$ are connected,} \\
         0, & \mbox{otherwise.}
     \end{array}
\right.
\end{equation}

This model has been studied by means of Monte Carlo simulations,
sampling the configuration space by using the Metropolis
local update algorithm \cite{bi10}. 
We are particularly interested in the behavior of the magnetization 
$M = \sum_{i=1}^N S_i/N$. For a given set of parameters
($\gamma$, $\nu$, $k_0$) defining the networks, the average value 
$\langle M \rangle$ has been studied as a function of temperature 
$T$ and system size $N$.  This allows us to investigate the transition 
from a FM ($\langle M \rangle \neq 0$) 
to a PM ($\langle M \rangle$ = 0) regime as $T$ is increased.

Depending on the value of the exponent $\gamma$ defining the power-law 
distribution of single edges, two different cases are 
found \cite{le02,do02b,he04}. 
First, for $\gamma > 3$, one expects a phase transition with
a well-defined transition temperature $T_c$ ($<\infty$) in the thermodynamic 
limit $N \to \infty$.
Second, for $\gamma \leq 3$ a FM--PM crossover is known to appear for
scale-free networks, with a crossover temperature $T_{\rm co}(N)$ increasing 
with system size and diverging to infinity as $N \to \infty$.

For the cases where a FM--PM transition occurs in the thermodynamic limit
($\gamma > 3$), the transition temperature $T_c$ has been obtained here by 
using Binder's fourth-order cumulant \cite{bi10}
\begin{equation} \label{Binder}
 U_N(T) \equiv 1 - \frac{ \langle M^4 \rangle_N } {3 \langle M^2 \rangle^2_N} 
   \, ,
\end{equation} 
The average values in this expression are taken over different network 
realizations and 
over different spin configurations for a given network at temperature $T$.  
In this case, the transition temperature is obtained from the unique crossing 
point of the functions $U_N(T)$ for several system sizes $N$ \cite{he02,he04}. 

In the second case ($\gamma \leq 3$), the size-dependent crossover temperature 
$T_{\rm co}(N)$ has been obtained from the maximum of the magnetization 
fluctuations $(\Delta M)^2_N$ as a function of temperature, with 
\begin{equation}
  (\Delta M)^2_N = \langle M^2 \rangle_N - \langle M \rangle^2_N   \, . 
\end{equation}
We note that $T_{\rm co}$ values derived by using this criterion agree 
within error bars with those found from the maximum 
derivative of the heat capacity \cite{he04}. 

The largest networks considered here included about $10^5$ sites.  
Such network sizes were employed in particular to study the dependence of
$T_{\rm co}$ on $N$ for $\gamma \leq 3$.
For the cases where a phase transition exists in the thermodynamic limit 
($\gamma > 3$), sizes around $4 \times 10^4$ nodes were considered.
The results presented below were obtained by averaging in each case over 
800 networks, except for the largest system sizes, for which 400 network
realizations were generated. 
Similar MC simulations have been carried out earlier to study
ferromagnetic \cite{he02,he04} and antiferromagnetic \cite{he08,he09} 
Ising models in complex networks.

\section{Results and discussion}

\subsection{Case $\gamma > 3$}

For unclustered scale-free networks with an exponent $\gamma > 3$, 
the average value 
$\langle k^2 \rangle$ converges to a finite value as $N \to \infty$. 
In this case, analytical calculations \cite{le02,do02b} and Monte Carlo
simulations \cite{he04} predict a well-defined FM--PM transition 
temperature $T_c$ given by
\begin{equation} 
\frac{J}{T_c} = \frac{1}{2}  \ln \left( \frac {\langle k^2 \rangle}
        { \langle k^2 \rangle - 2 \langle k \rangle } \right)  \; .
\label{Tcan}
\end{equation} 
This equation is equivalent to Eq.~(56) in Ref.~\onlinecite{yo11}.

For the clustered networks considered here with $\gamma > 3$, 
we have calculated $T_c$ from the Binder's cumulant $U_N$ for 
several values of the triangle 
density $\nu$ and minimum degree $k_0 > 1$.
In each case, four different network sizes were considered.
In all these cases, a well defined transition temperature was found
from the crossing point of the curves $U_N(T)$ for different sizes $N$,
as in Ref.~\onlinecite{he04}.
We note that $\langle k^2 \rangle - 2 \langle k \rangle > 0$
for $k_0 > 1$, and $T_c$ is well defined by Eq.~(\ref{Tcan}) .  
For $k_0 = 1$, the simulated networks consist of many 
disconnected components, and Binder's cumulant $U_N(T)$ does not give a 
unique crossing point for different system sizes $N$.

\begin{figure}
\vspace{-1.0cm}
\includegraphics[width= 8cm]{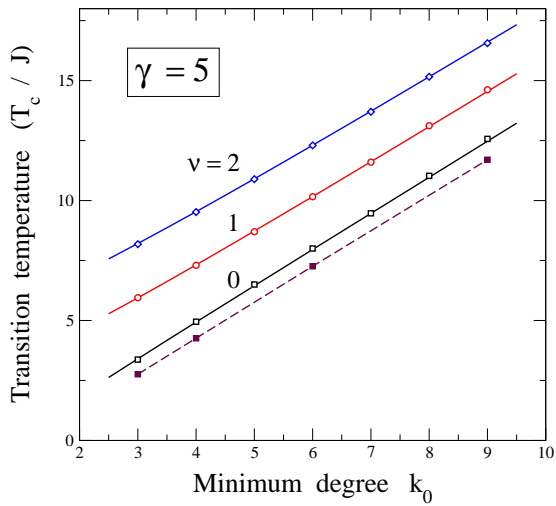}
\vspace{-0.3cm}
\caption{
Transition temperature $T_c$ for networks with $\gamma = 5$ as a
function of the minimum degree $k_0$.
Open symbols represent results of MC simulations, as obtained from
Binder's cumulant for $\nu$ = 0 (squares), 1 (circles), and 2 (diamonds).
Error bars are less than the symbol size.
Solid lines were derived from the analytical expression given in
Eq.~(\ref{Tcan}), using  values of $\langle k \rangle$ and
$\langle k^2 \rangle$ obtained from Eqs.~(\ref{kmean}) and (\ref{k2mean}).
Solid squares are data obtained in Ref.~\onlinecite{he04} for unclustered
networks ($\nu$ = 0).  The dashed line is a guide to the eye.
}
\label{f4}
\end{figure}

Going to the results of the present MC simulations,
in Fig.~4 we present the transition temperature $T_c$ as a function of the
minimum degree $k_0$ for an exponent $\gamma = 5$, and three values of
the triangle density $\nu$ = 0, 1, and 2.
In the three cases we observe a linear increase of $T_c$ for rising $k_0$.
The case $\nu = 0$ corresponds to a power-law degree distribution 
$P_{\rm sf}(k) \sim k^{-\gamma}$ (unclustered networks). 
Values of $T_c$ found here for these networks are 
somewhat higher than those obtained in Ref.~\onlinecite{he04}, by
an amount $\Delta T_c \sim 0.6 \, J$, due to the strict degree cutoff 
$k_{\rm cut}$ employed in that work.
These earlier results are shown in Fig.~4 as open symbols.
For $\nu = 1$ we find $T_c$ values higher than for $\nu = 0$, and the 
transition temperature increases further, by the same amount, for
$\nu = 2$.

\begin{figure}
\vspace{-1.0cm}
\includegraphics[width= 8cm]{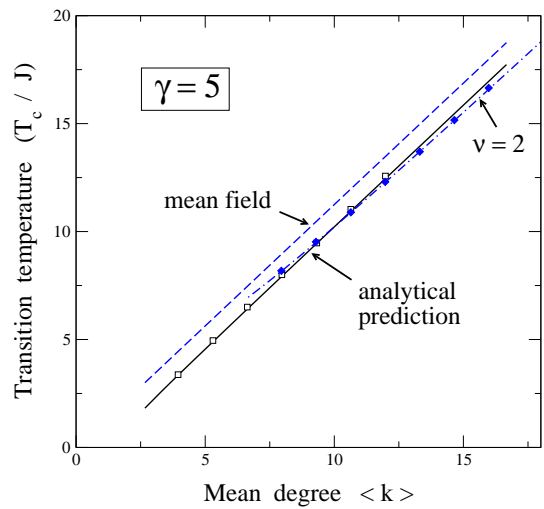}
\vspace{-0.3cm}
\caption{
Transition temperature $T_c$ for scale-free networks with $\gamma = 5$ as a
function of the mean degree $\langle k \rangle$.
Data points are results for networks with $\nu$ = 0 (open squares)
and $\nu$ = 2 (solid diamonds). Error bars are less than the symbol size.
Solid and dashed-dotted lines are analytical predictions from
Eq.~(\ref{Tcan}) for $\nu$ = 0 and 2, respectively.
The dashed line represents the mean-field result given by Eq.~(\ref{Tcmf}).
}
\label{f5}
\end{figure}

Looking at Eq.~(\ref{Tcan}), one expects that $T_c$ should have an explicit
dependence on $\langle k \rangle$ and $\langle k^2 \rangle$, rather than
the minimum degree $k_0$ itself.
In Fig.~5 we present the transition temperature $T_c$ as a function of 
the mean degree $\langle k \rangle$, as derived from MC simulations
for unclustered networks ($\nu = 0$, open squares) and clustered networks
with $\nu = 2$ (solid diamonds).  Lines in this figure were obtained 
from the average values $\langle k \rangle$ and $\langle k^2 \rangle$.
Taking into account that $k_i = s_i + 2 t_i$, the average value 
$\langle k^2 \rangle$ for clustered networks with $\gamma > 3$ 
can be calculated as:
\begin{equation}
    \langle k^2 \rangle = \langle s^2 \rangle + 
           4 \langle s \rangle \langle t \rangle +
           4 \langle t^2 \rangle   \; ,
\end{equation}
since $s$ and $t$ are independent due to the way of building up 
these networks. 
For the power-law distribution of $s_i$ corresponding to single edges,
we have in the large-$N$ limit:
\begin{equation}
    \langle s^2 \, \rangle_{\infty} 
        \approx k_0^2 \, \frac{\gamma - 1}{\gamma - 3}
\end{equation}
so that 
\begin{equation}
 \langle k^2 \rangle_{\infty} = k_0^2 \, \frac{\gamma - 1}{\gamma - 3} +
        4 \, k_0 \, \nu \, \frac{\gamma - 1}{\gamma - 2} + 
        4 \, \nu \, (\nu + 1)      \, .
\label{k2mean}
\end{equation}
Here $\nu$ and $\nu \, (\nu + 1)$ are the average values 
$\langle t \rangle$ and $\langle t^2 \rangle$ corresponding to the
Poisson distribution $Q(t)$ of triangles in these networks.
Introducing Eq.~(\ref{k2mean}) for $\langle k^2 \rangle_{\infty}$ and 
Eq.~(\ref{kmean}) for $\langle k \rangle_{\infty}$ into Eq.~(\ref{Tcan}), 
with the parameters  $\gamma = 5$ and $\nu = 0$,
we find for the transition temperature $T_c$ the solid line shown 
in Fig.~5.  This line lies very close to the results derived from 
our Monte Carlo simulations for unclustered networks (open squares). 
Similarly, for $\gamma = 5$ and $\nu = 2$, we obtain from 
Eq.~(\ref{Tcan}) the dashed-dotted line, which coincides with the data
obtained from simulations for clustered networks (full diamonds).
Note that for $\gamma > 3$, finite-size effects on 
$\langle k \rangle$ and $\langle k^2 \rangle$ are negligible for the network
sizes employed in our simulations, so that the agreement between 
Eq.~(\ref{Tcan}) and our simulation data is good. In fact, the MC results 
agree within error bars with the transition temperature given by
Eq.\,(\ref{Tcan}).

For comparison, we also present in Fig.~5 the critical temperature obtained 
in a mean-field approach \cite{le02}: 
\begin{equation}
  T_c^{MF} = \frac{\langle k^2 \rangle}{\langle k \rangle} \, J   \, ,
\label{Tcmf}
\end{equation}
which is displayed as a dashed line ($\nu = 0$). Note that this mean-field 
expression can be derived from Eq.~(\ref{Tcan}) in the limit
$\langle k^2 \rangle / \langle k \rangle \gg 1$.
Expanding Eq.~(\ref{Tcan}) for small 
$\langle k \rangle / \langle k^2 \rangle$, one has
\begin{equation}
  \frac{T_c}{J} = \frac{\langle k^2 \rangle}{\langle k \rangle} - 1 -
         \frac13 \frac{\langle k \rangle}{\langle k^2 \rangle} +
         {\cal O} \left( \frac{\langle k \rangle^2}{\langle k^2 \rangle^2} 
         \right)
\label{tcj_ser}
\end{equation}
where we recognize the first term in the expansion as the mean-field
approximation in Eq.~(\ref{Tcmf}).

The critical temperature $T_c$ derived from our MC simulations for
$\gamma = 5$, and shown in Fig.~4 as a function of the minimum degree
$k_0$, can be fitted linearly with good precision as 
$T_c = a \, k_0 + b$. In fact, for $\nu = 0$ we find $a = 1.52$ and 
$b = -1.17$.  
The value of $a$ can be estimated from the mean-field approximation
in Eq.~(\ref{Tcmf}), which yields $T_c \approx 3 k_0 J / 2$.

Turning to the results found for clustered networks, we observe in
Fig.~4 that, for a given $k_0$, the transition temperature clearly 
increases when the triangle density rises.
However, the same expression for $T_c$ given in Eq.~(\ref{Tcan})
reproduces well the MC results for clustered and unclustered networks,
once the corresponding values for $\langle k \rangle$ and 
$\langle k^2 \rangle$ are introduced, as shown in Fig.~5.
Since these average values depend only on the degree distribution
$P(k)$ (nearest neighbors), this means that the transition temperature
for networks with $\gamma > 3$ does not depend on the clustering.
Thus, including triangles in these networks changes the transition 
temperature because it changes the degree distribution $P(k)$, but 
networks with the same $P(k)$ but without triangles give the same
$T_c$, as predicted by Eq.~(\ref{Tcan}).  This does not happen for the 
crossover temperature $T_{\rm co}$ obtained for finite-size networks
with $\gamma \leq 3$ (see below).

\subsection{Case $\gamma = 3$}

For unclustered scale-free networks with an exponent $\gamma$ close to, 
but higher than 3, 
the transition temperature $T_c$ is given from Eq.~(\ref{tcj_ser}) by
\begin{equation}
\frac{T_c}{J} \approx \frac{\langle k^2 \rangle}{\langle k \rangle} 
          \approx  k_0 \, \frac{\gamma - 2}{\gamma -3}    \, .
\label{Tcj}
\end{equation}
Then, $d T_c / d \gamma  \approx  - k_0 J / (\gamma -3)^2 < 0$, and $T_c$ 
increases as $\gamma$ is reduced, eventually diverging for $\gamma \to 3$, 
as a consequence of the divergence of $\langle k^2 \rangle$.

For $\gamma = 3$, analytical calculations \cite{do02b,ig02} have predicted
a FM--PM crossover at a size-dependent temperature $T_{\rm co}$, which scales 
as $\log N$.  This dependence of the crossover temperature agrees with
that derived from MC simulations for the same type of networks \cite{he04}.
A logarithmic increase of $T_{\rm co}$ with system size $N$ has been also 
found by Aleksiejuk {\em et al.} \cite{al02b,al02c} from MC simulations 
of the Ising model in Barab\'asi-Albert growing networks. 
Note that these networks (with $\gamma = 3$) display correlations between 
degrees of adjacent nodes \cite{ba99b}.

For scale-free networks with $\gamma = 3$ and $\nu = 0$, the mean degree
$\langle k \rangle$ can be approximated as   [see Eqs.~(\ref{kmean})  
and (\ref{kmean3})]:
\begin{equation}
  \langle k \rangle \approx  2 k_0
     \left[ 1 - \left( \frac{c}{N} \right)^{\frac12} \right]  \;  ,
\end{equation}
and $\langle k^2 \rangle$ is given by [see Appendix, Eq.~(\ref{k2mean3})]:
\begin{equation}
    \langle k^2 \rangle  \approx k_0^2 \, \ln N  \;  .
\end{equation}
Applying Eq.~(\ref{Tcan}) to the size-dependent crossover temperature
corresponding to $\gamma = 3$, one finds for large system size $N$:
\begin{equation}
    T_{\rm co}  \approx  \frac12  k_0 J  \ln N     \; .
\label{Tco}
\end{equation}
For uncorrelated scale-free networks with $\gamma = 3$,
Dorogovtsev {\em et al.} \cite{do02b} found for the crossover temperature
\begin{equation}
  T_{\rm co}  \approx \frac{1}{4} \langle k \rangle  J  \ln N   \, ,
\label{Tco2}
\end{equation}
which coincides with Eq.~(\ref{Tco}) for
$\langle k \rangle \approx k_0 (\gamma - 1) / (\gamma - 2) = 2 k_0$.

\begin{figure}
\vspace{-1.0cm}
\includegraphics[width= 8cm]{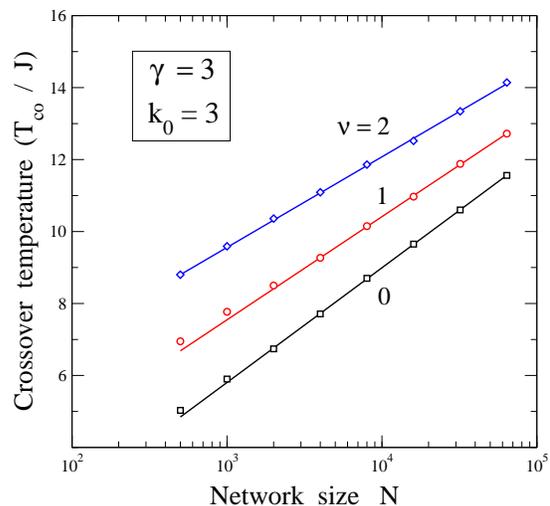}
\vspace{-0.3cm}
\caption{
Crossover temperature $T_{\rm co}/J$ for scale-free networks with $\gamma = 3$
and minimum degree $k_0 = 3$, as a function of the system size $N$.
Squares, $\nu = 0$; circles, $\nu = 1$; diamonds, $\nu = 2$.
Lines are least-square fits to the data points for $N \geq 4000$ nodes.
}
\label{f6}
\end{figure}

In Fig.~6 we present the results of our MC simulations for $T_{\rm co}$
as a function of the network size $N$. Data points correspond to 
networks with $\gamma = 3$ and three values of the triangle density
$\nu$ = 0, 1, and 2. 
The observed linear trend of the data points in this semilogarithmic
plot indicates a dependence $T_{\rm co} \sim J \ln N$, as that given
in Eq.~(\ref{Tco}).
We find that values of the crossover temperature for system size 
$N < 4000$ tend to be higher than the linear asymptotic trend 
found for larger sizes. In fact, in the linear fits presented in Fig.~6,
we only included sizes $N \geq 4000$. The deviation for small $N$ is
particularly observed for $\nu$ = 0 and 1.
Thus, our results indicate a dependence $T_{\rm co}/J = A \ln N + B$, 
with a constant $A$ that decreases for increasing $\nu$.
We found $A$ = 1.38, 1.25, and 1.09 for $\nu$ = 0, 1, and 2, respectively.
Note that in the case $\nu$ = 0 (scale-free networks without clustering),
the slope $A$ is somewhat smaller than the value predicted by Eq.~(\ref{Tco})
for $k_0 = 3$, i.e., $A = k_0 / 2 = 1.5$.

A similar logarithmic dependence of $T_{\rm co}$ upon $N$ was observed in 
earlier works.
For scale-free networks with a strict cutoff $k_{\rm cut}$, 
the prefactor $A$ was found to increase linearly with $k_0$, so that 
$A/k_0 = 0.28 \pm 0.01$ \cite{he04}.
For Barab\'asi-Albert networks with $k_0 = 5$, 
Aleksiejuk {\em et al.} \cite{al02b}
found from a fit similar to ours $A = 2.6$, which means $A/k_0 = 0.52$,
similar to our $A/k_0 = 0.46$ for $\nu = 0$.

\begin{figure}
\vspace{-1.0cm}
\includegraphics[width=8cm]{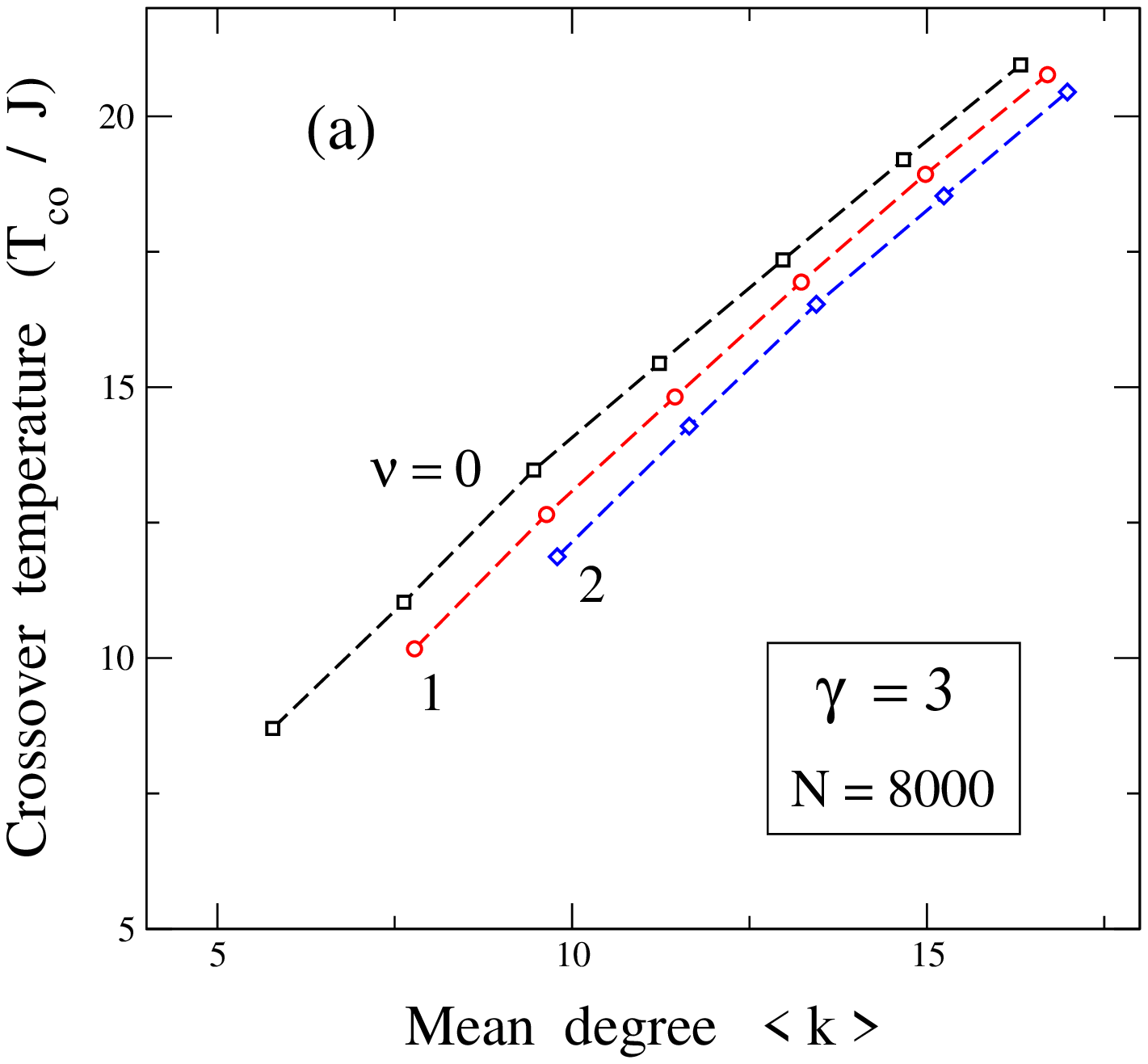}
\includegraphics[width=8cm]{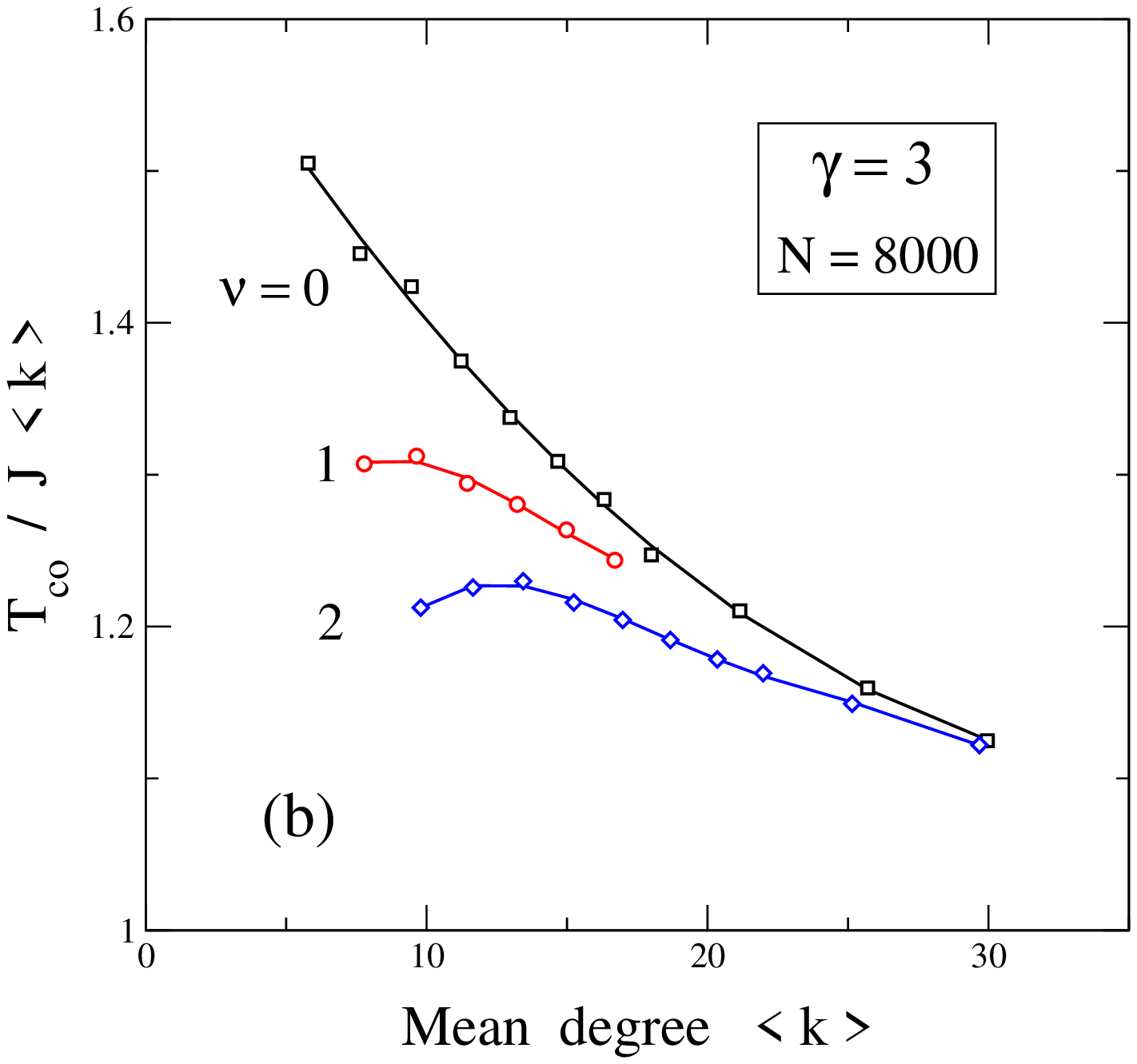}
\vspace{-0.3cm}
\caption{
(a) Crossover temperature as a function of the mean degree
$\langle k \rangle$, for networks with $\gamma = 3$ and
$N$ = 8000 nodes. Symbols represent results for three
$\nu$ values: 0 (squares), 1 (circles), and 2 (diamonds).
The data shown were obtained for networks with several
minimum degrees, $k_0 \geq 3$.
(b) Ratio $T_{\rm co} / J \langle k \rangle$ vs $\langle k \rangle$
for the same kind of networks as in (a).
Lines are guides to the eye.
}
\label{f7}
\end{figure}

Given the increase in crossover temperature for rising system size $N$, 
it is worthwhile analyzing the dependence of $T_{\rm co}$ on the minimum 
degree $k_0$. In fact, given the parameters $\gamma$ and $\nu$, $k_0$
controls the mean degree $\langle k \rangle$ of our networks.
In Fig.~7(a) we display results for $T_{\rm co}/J$ for an exponent 
$\gamma = 3$ and triangle density $\nu$ = 0, 1, and 2. In all cases, 
the networks included $N$ = 8000 nodes.
As expected, for a given value of $\nu$, the crossover temperature 
increases as $\langle k \rangle$ (or $k_0$) is raised.  Moreover, 
the line giving the dependence of $T_{\rm co}$ on $\langle k \rangle$
shifts downwards for rising $\nu$.
In view of Eq.~(\ref{Tcj}), this can be interpreted from a decrease in
the ratio $\langle k^2 \rangle / \langle k \rangle$ for networks with
constant size and increasing triangle density $\nu$.
Note, however, that the dependence of $T_{\rm co}$ on 
$\langle k \rangle$ is not strictly linear for fixed $N$, as predicted 
by Eq.~(\ref{Tco2}). This is a finite-size effect, since this equation 
corresponds to the asymptotic limit, valid in the large-$N$ regime,
so that for $N$ = 8000 such an effect is still clearly appreciable
in the results shown in Fig.~7(a).
This can be further visualized in Fig.~7(b), where we present 
the ratio $T_{\rm co} / (J \langle k \rangle)$ for the same data as 
in panel (a). Values of this ratio corresponding to different triangle
densities $\nu$ converge one to the other as the mean degree increases.

\begin{figure}
\vspace{-1.0cm}
\includegraphics[width= 8cm]{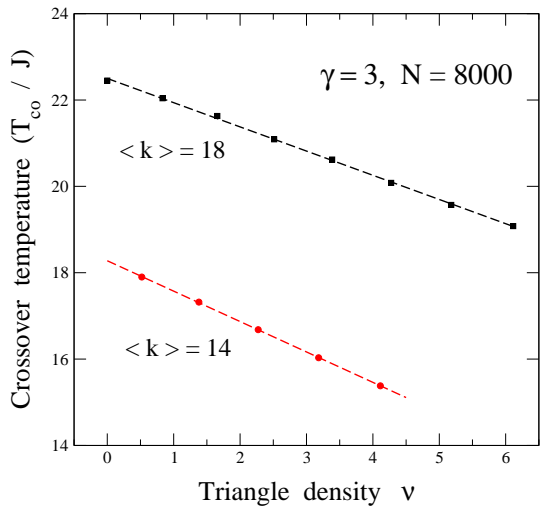}
\vspace{-0.3cm}
\caption{
Crossover temperature $T_{\rm co}/J$ for scale-free networks with $\gamma = 3$
and $N$ = 8000 nodes, as a function of the triangle density $\nu$.
Each set of data corresponds to a fixed value of the mean degree:
$\langle k \rangle$ = 14 (circles) and $\langle k \rangle$ = 18 (squares).
Dashed lines are least-square fits to the data points.
}
\label{f8}
\end{figure}

For networks with a given size $N$, it is interesting to analyze
the dependence of $T_{\rm co}$ on the parameter $\nu$ for a fixed value
of the mean degree $\langle k \rangle$.
Since $\langle k \rangle = \langle s \rangle + 2 \nu$, one can obtain
networks with given $\langle k \rangle$ by simultaneously changing
$\langle s \rangle$ and $\nu$. In the actual implementations, we
varied the minimum degree $k_0$, which defines a mean value
$\langle s \rangle$, and then we took 
$\nu = (\langle k \rangle - \langle s \rangle) / 2$, $\langle k \rangle$ 
being the required mean degree of the clustered networks.   
In Fig.~8 we show the dependence of $T_{\rm co}$ on the triangle density
$\nu$ for $\langle k \rangle$ = 14 and 18.
As expected from the data shown in Fig.~7, $T_{\rm co}$ decreases
for rising $\nu$, and this dependence turns out to be linear in both
cases considered here.
The slope of this line is more negative for $\langle k \rangle$ = 14
than for $\langle k \rangle$ = 18. In fact, we found:
$d T_{\rm co} / d \nu$ = --0.70 $J$ and --0.56 $J$ for $\langle k \rangle$ = 
14 and 18, respectively.

This means that for networks with given size $N$ and mean degree 
$\langle k \rangle$, including triangles in the networks (i.e., increasing
the triangle density $\nu$) reduces the crossover temperature $T_{\rm co}$.
The reason for this is the following. To have a constant 
$\langle k \rangle$ when changing $\nu$ for a given $\gamma$
($\gamma = 3$ here), one needs a minimum degree
$k_0 = \langle k \rangle / 2 - \nu$ [see Eq.~(\ref{kmean})], 
so that a rise of $\nu$ is associated to a decrease in $k_0$.
This causes a reduction of $\langle k^2 \rangle$, and therefore
the temperature $T_{\rm co}$ decreases.

In other words, adding triangles to a network changes the degree 
distribution $P(k)$ itself, apart from introducing clustering into the
network. When one includes triangles ($\nu$ increases) without 
changing the other parameters defining the networks 
(i.e., $N$, $k_0$, $\gamma$),
one finds a clear increase in $T_{\rm co}$, as shown in Fig.~6.
However, if one changes $\nu$ subject to some particular restriction on
the degree distribution, such as keeping constant the mean value
$\langle k \rangle$, one may find other types of dependence of 
$T_{\rm co}$ on $\nu$ (as the decrease shown in Fig.~8).

This suggests that a relevant point here is
a comparison between clustered and unclustered networks with the
same degree distribution $P(k)$. This will give insight into the
`direct' effect of clustering on the critical properties of the Ising 
model.    As commented above, the
distribution $P(k)$, as well as Eq.~(\ref{Tcan}) predicting the crossover 
temperature, do not include any information on the clustering present in 
the considered networks, but only on the degrees (connectivity) of the nodes.
Thus, a natural question is the relevance of the difference between the
crossover temperature corresponding to networks with the same size $N$ and 
degree distribution $P(k)$, but including triangles or not.
We have seen above that in the case $\gamma > 3$ both types of networks
have the same transition temperature $T_c$, which agrees with 
Eq.~(\ref{Tcan}). This is not clear, however, for $\gamma = 3$.
To clarify this point we have generated networks with the same $P(k)$
as those studied above for different $\nu$ values, but without 
including triangles.
In this case, we used the distribution $P(k)$ to define the set of degrees
$\{ s_i  \}_{i=1}^N$, employed to build up the networks
($k_i = s_i$ for all $i$).

\begin{figure}
\vspace{-1.0cm}
\includegraphics[width= 8cm]{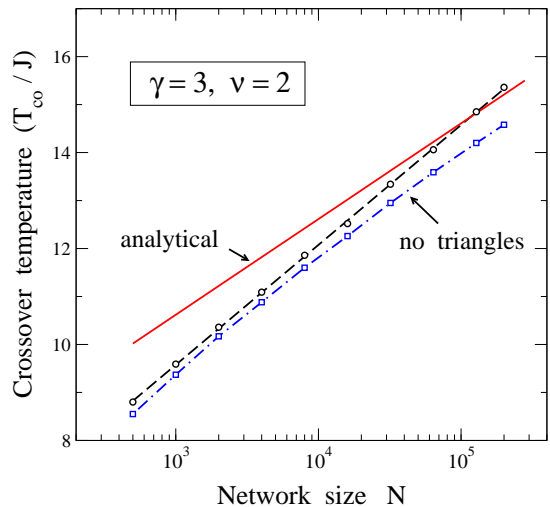}
\vspace{-0.3cm}
\caption{
Crossover temperature as a function of system size $N$ for
networks with $\gamma = 3$, $\nu = 2$, and minimum degree $k_0 = 3$.
Circles and squares represent results of MC simulation for networks
with and without triangles, but with the same degree distribution $P(k)$.
Error bars are less than the symbol size.
The solid line is the analytical prediction, Eq.~(\ref{Tcan}), obtained
from the mean values $\langle k \rangle$ and $\langle k^2 \rangle$.
}
\label{f9}
\end{figure}

In Fig.~9 we present the temperature $T_{\rm co}$ as a function of 
system size for networks with (circles) and without (squares) triangles,
as derived from our MC simulations. Clustered networks were generated with
$\gamma = 3$, $\nu = 2$, minimum degree $k_0 = 3$, and several sizes
up to $N = 2 \times 10^5$ nodes. Unclustered networks were built up
with the same distribution $P(k)$ as the clustered ones.
The results presented in Fig.~9, indicate first that clustering increases
$T_{\rm co}$, and this increase becomes more important for larger system size.
As discussed above, for the clustered networks we find a dependence 
$T_{\rm co}/J = A \ln N + B$, with a slope $A$ close to unity ($A = 1.09$).
We have also plotted the crossover temperature predicted by
Eq.~(\ref{Tcan}), which takes only into account the average values
$\langle k \rangle$ and $\langle k^2 \rangle$ (solid line).
A linear dependence on $\ln N$ is also found in this case, 
but with a smaller coefficient $A = 0.87$.
In fact, this line crosses with that derived from MC simulations
for a system size $N \approx 10^5$.

Looking at the size dependence of the crossover temperature for
unclustered networks shown in Fig.~9 (dashed-dotted line), we observe 
that for $N < 10^4$ the curve $T_{\rm co}(N)$ found for these networks 
is parallel (slightly below) to that found for clustered networks
(dashed line).  For larger system sizes, the dashed-dotted curve becomes 
parallel to that corresponding to the analytical model (solid line).

For a given system size, clustered networks display a crossover temperature
$T_{\rm co}$ larger than unclustered networks with the same degree
distribution $P(k)$. This means that clustering (triangles in this case)
favors an increase in $T_{\rm co}$, i.e. the FM phase is stable in a
broader temperature range.
This has been already observed in the results shown in Fig.~6, but in that
case the difference between clustered and unclustered networks was
larger, due to the inclusion of triangles for $\nu > 0$, which changed
the actual degree distribution $P(k)$ with respect to the case $\nu = 0$.

The result for unclustered networks (no triangles) shown in Fig.~9
converges to the analytical data given by $T_{\rm co} \sim A J \ln N$,
with $A$ = 0.87, as expected from the asymptotic limit for Eq.~(\ref{Tcan}).
For large system size, the behavior of unclustered networks is in this
respect controlled by nodes with large degree. Given that the effective
degree cutoff scales as $k_{\rm cut} \sim N^{1/2}$ [see Eq.~(\ref{kcutk})],
nodes with large $k$ appear progressively as $N$ is increased.
Thus, the presence of nodes with high degree favors ferromagnetic
correlations in clustered networks, and therefore an increase in 
the crossover temperature $T_{\rm co}$.

\begin{figure}
\vspace{-1.0cm}
\includegraphics[width= 8cm]{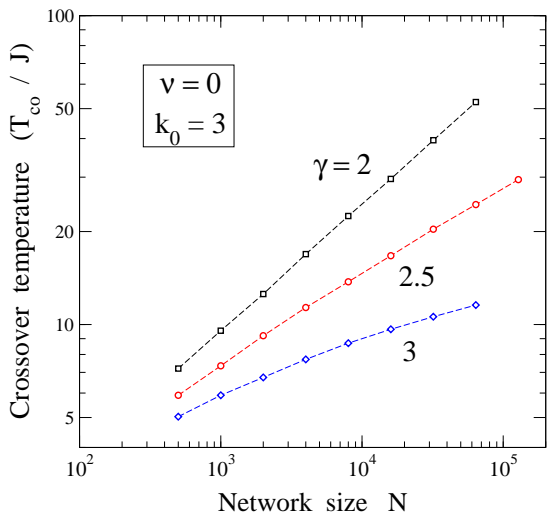}
\vspace{-0.3cm}
\caption{
Crossover temperature $T_{\rm co}/J$ for scale-free networks with $\nu = 0$
and $k_0 = 3$, as a function of system size $N$ in a logarithmic plot.
Symbols represent results for three values of the parameter $\gamma$:
2 (squares), 2.5 (circles), and 3 (diamonds).
Error bars are less than the symbol size.
Lines are guides to the eye.
}
\label{f10}
\end{figure}

\subsection{Case $\gamma < 3$}

As commented above, for $\gamma < 3$ it is known that correlations between 
degrees of adjacent nodes appear in scale-free networks when no multiple 
and self-connections are allowed, unless one takes a
degree cutoff $k_{\rm cut} \lesssim \sqrt{N}$ \cite{ca05b}. 
For this reason, we generated clustered and unclustered networks with 
$\gamma < 3$ assuming a cutoff $k_{\rm cut} = \sqrt{N}$.
This means that in this case Eq.~(\ref{sumkc}) does not apply.
With a calculation similar to that presented in the Appendix, one
finds in this case for unclustered scale-free networks ($\nu = 0$):
\begin{equation} 
 \langle k^2 \rangle \approx \frac{\gamma-1}{3-\gamma} \, k_0^{\gamma-1}
      \,   N^{(3-\gamma)/2}  \; .
\end{equation} 
For $2 < \gamma < 3$, one has for the mean degree: 
\begin{equation}
   \langle k \rangle \approx k_0 \, \frac{\gamma-1}{\gamma - 2}   \; ,
\end{equation}
as in Eq.~(\ref{kmeaninf}). For $\gamma = 2$, $\langle k \rangle$ 
diverges to infinity in the large-size limit as 
\begin{equation}
   \langle k \rangle \approx \frac12 \, k_0 \, \ln N  \; .
\end{equation}
Thus, one expects a size-dependent crossover temperature 
\begin{equation}
 \frac{T_{\rm co}}{J} \approx \frac{\langle k^2 \rangle}{\langle k \rangle}
    \approx \frac{\gamma-2}{3-\gamma} \, k_0^{\gamma-2}  \,
            N^{(3-\gamma)/2}
\label{Tcoj}
\end{equation}
for $2 < \gamma < 3$, and
\begin{equation}
   \frac{T_{\rm co}}{J}  \approx  2 \, \frac{\sqrt{N}}{\ln N}
\end{equation}
for $\gamma = 2$.

\begin{figure}
\vspace{-1.0cm}
\includegraphics[width=8cm]{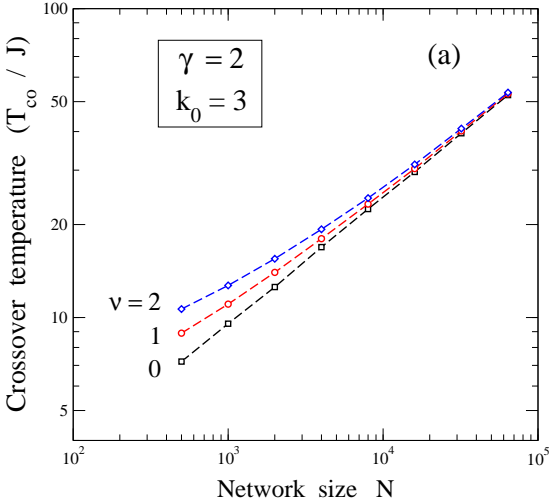}
\includegraphics[width=8cm]{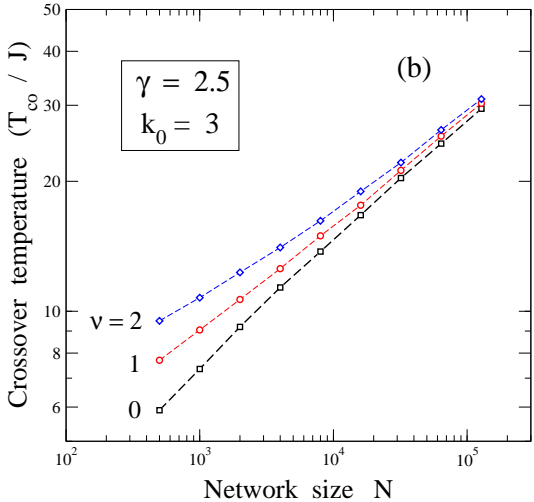}
\vspace{-0.3cm}
\caption{
Crossover temperature $T_{\rm co} / J$ as a function of system size $N$
for scale-free networks with minimum degree $k_0$ = 3, and three values
of the triangle density $\nu$: 0 (squares), 1 (circles), and 2 (diamonds).
(a) Networks with $\gamma = 2$; (b) networks with $\gamma = 2.5$.
Lines are guides to the eye.
}
\label{f11}
\end{figure}

We first present results for unclustered networks ($\nu = 0$).
In Fig.~10 we show the temperature $T_{\rm co}$ as a function of system 
size $N$ for three values of $\gamma$ in a logarithmic plot, as derived from 
our MC simulations for networks with minimum degree $k_0 = 3$. 
The exponent $\gamma$ increases from top to bottom: $\gamma$ = 2, 2.5, and 3.
For a given system size, $T_{\rm co}$ decreases as $\gamma$ rises.
This is a consequence of a decrease in the ratio 
$\langle k^2 \rangle / \langle k \rangle$ for increasing $\gamma$, which
causes a reduction in $T_{\rm co}$, as predicted by Eq.~(\ref{Tcoj}),
where the term $N^{(3-\gamma)/2}$ dominates for large $N$. 

 For networks with $\gamma < 3$ and large-enough size, $\log T_{\rm co}$ 
derived from the MC simulations is found to display a linear dependence 
on $\log N$, as expected for a crossover temperature 
diverging as a power of the system size $T_{\rm co} \sim N^z$ 
with an exponent $z$ dependent on the parameter $\gamma$ \cite{he04}. 
Such a linear dependence is obtained for system sizes $N \gtrsim N_0$, the 
size $N_0$ increasing with the exponent $\gamma$ and eventually diverging for 
$\gamma \to 3$, for which $T_{\rm co} \sim \ln N$ is expected (see Fig.~6). 
According to Eq.~(\ref{Tcoj}), one expects $z = (3-\gamma)/2$.
In fact, for $\gamma = 2.5$ we find $z = 0.27$ from a fit to the data points
corresponding to networks with $N > 10^4$. This value of $z$ could further
decrease for larger system sizes, what is compatible with the exponent
$z = 0.25$ expected from Eq.~(\ref{Tcoj}).

\begin{figure}
\vspace{-1.0cm}
\includegraphics[width= 8cm]{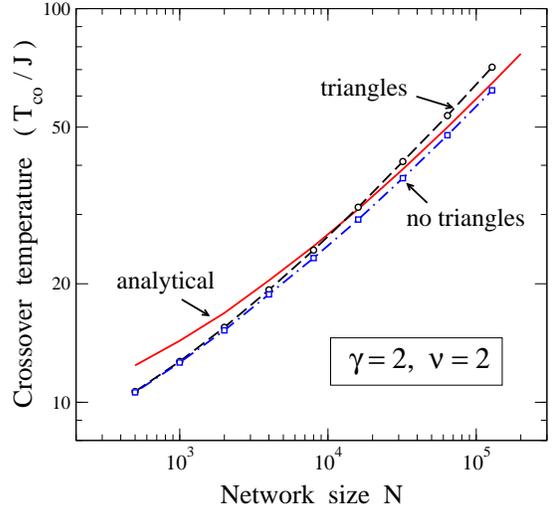}
\vspace{-0.3cm}
\caption{
Crossover temperature as a function of system size $N$ for
networks with $\gamma = 2$, $\nu = 2$, and minimum degree $k_0 = 3$.
Circles and squares represent results of MC simulations for networks
with and without triangles, respectively, but with the same degree
distribution $P(k)$.  Error bars are less than the symbol size.
The solid line indicates the result derived from $\langle k \rangle$
and $\langle k^2 \rangle$ by using Eq.~(\ref{Tcan}).
}
\label{f12}
\end{figure}

We note that the degree cutoff is relevant for the 
size-dependence of the crossover temperature. Thus, the cutoff employed
here for networks with $\gamma < 3$ ($k_{\rm cut} = \sqrt{N}$) yields 
an exponent $z = (3-\gamma)/2$,
to be compared with $z = (3-\gamma)/(\gamma-1)$, given by the ``natural'' 
cutoff in Eqs.~(\ref{sumkc}) and (\ref{kcutk}) 
[see Refs.~\onlinecite{do02b,ig02,he04}]. 
The latter cutoff is known to introduce undesired correlations in networks 
such as those considered here with $\gamma < 3$, as commented above.

We now turn to clustered networks.
In Fig.~11 we show the crossover temperature vs system size $N$
for networks with different triangle densities: 
$\nu$ = 0, 1, and 2. In panels (a) and (b) results are given for 
$\gamma$ = 2 and 2.5, respectively.
For small network size, $T_{\rm co}$ appears to be higher for larger 
parameter $\nu$ in both cases. This difference, however, decreases 
as $N$ is increased, and for each value of $\gamma$ the results for
different triangle densities $\nu$ converge one to the other.
Thus, in the logarithmic plots of Fig.~11 differences between the crossover
temperature for different $\nu$ values become irrelevant for system
size $N \sim 10^5$.

Similarly to the case $\gamma = 3$ presented in Sec.~IV.B, for $\gamma < 3$
it is also interesting to compare results for clustered and unclustered 
networks with the same degree distribution $P(k)$. 
In Fig.~12 we present the temperature $T_{\rm co}$ as a function of
system size for networks with (circles) and without (squares) triangles,
as derived from our MC simulations.  Clustered networks were generated 
with $\gamma = 2$, $\nu = 2$, minimum degree $k_0 = 3$, and various 
system sizes.
For $N < 2000$ nodes, results for clustered and unclustered networks 
coincide one with the other within statistical noise.
For $N > 2000$, both sets of data progressively separate one from the other,
so that the crossover temperature for clustered networks (including triangles) 
is larger than that corresponding to unclustered networks (no triangles).
The solid line in Fig.~12 is the analytical prediction obtained from
Eq.~(\ref{Tcan}) by introducing the mean values 
$\langle k \rangle$ and $\langle k^2 \rangle$ corresponding to the 
actual networks.
We observe something similar to the case of $\gamma = 3$ shown in 
Fig.~9. For large $N$, results of MC simulations for unclustered
networks approach the analytical expectancy, lying below the solid line,
whereas data for clustered networks become higher than the analytical 
prediction, and progressively deviate from the latter as system size $N$
is increased.

As for the case $\gamma = 3$, we conclude that clustering favors a
stabilization of the FM phase vs the PM one, hence increasing the 
crossover temperature $T_{\rm co}$. This becomes more noticeable 
for larger network size, where nodes with higher degree
progressively appear.

\section{Summary}

We have studied the FM-PM transition for the Ising model in clustered 
scale-free networks by means of Monte Carlo simulations, and the results
were compared with those found for unclustered networks.
Our results can be classified into two different regions, as a function of
the exponent $\gamma$ defining the power-law for the degree distribution
in scale-free networks. 

For $\gamma > 3$, we find in all cases a well-defined transition 
temperature $T_c$ in the thermodynamic limit, in agreement with earlier 
analytical calculations and MC simulations. This refers equally to 
clustered and unclustered networks.
Adding motifs (triangles in our case) to the networks causes an increase 
in the transition temperature, as a consequence of the associated change in 
the degree distribution $P(k)$, and in particular in the mean values
$\langle k \rangle$ and $\langle k^2 \rangle$.
However, for clustered and unclustered networks
with the same degree distribution $P(k)$, one finds no difference 
in $T_c$, which coincides with that predicted by analytical calculations
[Eq.~(\ref{Tcan})]. 
This conclusion agrees with that drawn in 
Ref.~\onlinecite{yo11}, where the addition of motifs was found to 
increase the transition temperature, without changing the critical
behavior. 

For networks with $\gamma \leq 3$, the situation is different.
In this case, the  crossover temperature $T_{\rm co}$ increases with system
size $N$.  For $\gamma = 3$ we found $T_{\rm co} \sim J \ln N$, and for
$\gamma < 3$ we obtained $T_{\rm co} \sim J N^z$, with an exponent 
$z = (3-\gamma)/2$. 
Comparing clustered and unclustered networks, the conclusions obtained
for $\gamma \leq 3$ differ from those found for $\gamma > 3$.

For $\gamma \leq 3$, $T_{\rm co}$ is similar for 
clustered and unclustered networks with the same degree distribution 
$P(k)$, when one considers small network sizes ($N \lesssim 10^3$).
This behavior changes for larger networks, and the crossover temperature
$T_{\rm co}$ of clustered networks becomes progressively larger than that
corresponding to the unclustered ones.
Thus, we find that FM correlations are favored by including triangles
in the networks, in particular in the presence of nodes with large
degree $k$.

It is important to note that this conclusion refers to a comparison of 
clustered and unclustered networks with the same degree distribution 
$P(k)$. Caution should be taken when comparing results for networks with 
different degree distributions. 
Thus, when triangles are added on a network with a given distribution
of links, the crossover temperature $T_{\rm co}$ increases 
[see Fig.~6], but in this case the rise in $T_{\rm co}$ is mainly
due to the change in $P(k)$ caused by the inclusion of triangles.
However, for clustered networks with the same exponent $\gamma$ 
(large-degree tail) and mean degree $\langle k \rangle$, an increase 
in triangle density $\nu$ causes a decrease in $T_{\rm co}$ [see Fig.~8],
as a consequence of the associated variation in the distribution 
$P(k)$, and in particular in the mean value $\langle k^2 \rangle$.

We also note that results for unclustered or clustered scale-free networks
may differ when different degree cutoffs are employed, especially for 
$\gamma < 3$, as the dependence of their properties on system size can be 
appreciable.  This applies, in particular, to the
scaling of the crossover temperature $T_{\rm co}$ on $N$ for large system 
size. To avoid correlations between degrees of adjacent nodes,
we have employed here for power-law distributions a degree cutoff
$k_{\rm cut} = N^{1/2}$. However, for clustered networks the presence 
of triangles introduces degree correlations. 

Other distributions different from the short-tailed Poisson-type 
introduced here for the triangles, could be considered to 
change more dramatically the long-degree tail of the overall degree
distribution $P(k)$. For example, a power-law distribution for
the triangles (with an exponent $\gamma'$) may give rise to an interesting 
competition between the exponents of both distributions (for single links
and triangles), which could change the critical behavior of the
Ising model on such networks as compared to those discussed here.

\begin{acknowledgments}
This work was supported by Direcci\'on General de Investigaci\'on (Spain)
through Grant FIS2012-31713,
and by Comunidad Aut\'onoma de Madrid
through Program MODELICO-CM/S2009ESP-1691.
\end{acknowledgments}

\appendix

\section{Finite-size effects on degree distributions}

Here we present some expressions related to finite-size effects
in scale-free networks.
We assume a degree distribution $P_{\rm sf}(k)$ defined as
\begin{equation}
P_{\rm sf}(k) =  \left\{
     \begin{array}{ll}
         n \, k^{-\gamma}, & {\rm for} \, k \geq k_0 \\
         0, & {\rm for} \, k < k_0
     \end{array}
\right.
\label{psf}
\end{equation}
with $\gamma > 2$ and $n$ a normalization constant.

Due to the finite network size $N$, an effective cutoff $k_{\rm cut}$ 
appears for the degree distribution in these networks \cite{do02b,ig02}. 
This cutoff is such that
$\sum_{k_{\rm cut}}^{\infty} P_{\rm sf}(k) \sim 1/N$, indicating that 
the number of nodes with $k > k_{\rm cut}$ is expected to be on the order 
of unity.  For concreteness, we write
\begin{equation}
\sum_{k_{\rm cut}}^{\infty} P_{\rm sf}(k) = \frac{c}{N}
\label{sumkc2}
\end{equation}
with $c = {\cal O}(1)$.
Replacing the sum by an integral, we find
\begin{equation}
    \sum_{k_{\rm cut}}^{\infty} P_{\rm sf}(k) \approx
           \frac{n}{\gamma-1} \, k_{\rm cut}^{1-\gamma}  \; .
\label{sumkc3}
\end{equation}
For $c/N \ll 1$, one has for the normalization condition:
\begin{equation}
  1 = \frac{c}{N} + \sum_{k_0}^{k_{\rm cut}} P_{\rm sf}(k) \approx
      \frac{n}{\gamma - 1} \, ( k_0^{1-\gamma} - k_{\rm cut}^{1-\gamma} )     
      \; .
\end{equation}
Then, the normalization constant $n$ is:
\begin{equation}
    n = \frac {\gamma - 1} {k_0^{1-\gamma} - k_{\rm cut}^{1-\gamma}}
        \approx (\gamma-1) \, k_0^{\gamma-1}  \, .
\label{nn}
\end{equation}

  Combining Eqs.~(\ref{sumkc2}), (\ref{sumkc3}), and (\ref{nn}) 
one finds
\begin{equation}
\frac{N}{c} = \left( \frac{k_{\rm cut}}{k_0} \right)^{\gamma-1} - 1  \; ,
\end{equation}
and for $k_{\rm cut} \gg k_0$:
\begin{equation}
 k_{\rm cut} \approx k_0 \left( \frac{N}{c} \right)^{\frac{1}{\gamma-1}} \, ,
\label{kcutk}
\end{equation}
so that $k_{\rm cut} \sim \, N^{1/(\gamma-1)}$, as in 
Refs.~\onlinecite{do02b,ig02}.
Considering this cutoff, one has for the mean degree
\begin{equation}
  \langle k \rangle = \sum_{k_0}^{k_{\rm cut}}  k \, P_{\rm sf}(k) =
     \frac{\gamma-1}{\gamma-2}  \, \,   
     \frac{ k_{\rm cut}^{2-\gamma} - k_0^{2-\gamma} }
          { k_{\rm cut}^{1-\gamma} - k_0^{1-\gamma} }
\end{equation}
which gives, using Eq.~(\ref{kcutk}):
\begin{equation}
  \langle k \rangle   \approx   \langle k \rangle_{\infty}  \,
     \frac  { \left( \frac{c}{N} \right)^{\frac{\gamma-2}{\gamma-1}} - 1 }
            {  \frac{c}{N} - 1 }  
\end{equation}
with $\langle k \rangle_{\infty} \approx k_0 (\gamma-1) / (\gamma-2)$.
Thus, we find for the mean degree $\langle k \rangle$:
\begin{equation}
     \langle k \rangle \approx \langle k \rangle_{\infty}
         \left[ 1 - \left( \frac{c}{N} \right)^{\frac{\gamma-2}{\gamma-1}}
            + {\cal O} \left( \frac1N \right)   \right]   \; .
\label{kmean_app}
\end{equation}

A similar calculation can be carried out to estimate finite-size effects on
$\langle k^2 \rangle$ for scale-free networks. 
For $\gamma > 3$, we find:
\begin{equation}
   \langle k^2 \rangle \approx \langle k^2 \rangle_{\infty}
        \left[ 1 - \left( \frac{c}{N} \right)^{\frac{\gamma-3}{\gamma-1}}
          + {\cal O} \left( \frac1N \right)   \right]
\end{equation}
with $\langle k^2 \rangle_{\infty} \approx k_0^2 \, (\gamma-1) / (\gamma-3)$.

For $\gamma = 3$, we have
\begin{equation}
    \langle k^2 \rangle =  n \sum_{k_0}^{k_{\rm cut}} \frac1k 
       \approx  \frac{2}{k_0^{-2} - k_{\rm cut}^{-2}}  \,
               \ln \frac{k_{\rm cut}}{k_0}
\end{equation}
and using Eq.~(\ref{kcutk}), we obtain for $k_{\rm cut} \gg k_0$:
\begin{equation}
  \langle k^2 \rangle =  k_0^2 \, \ln N + {\cal O} \left( 1 \right)
\label{k2mean3}
\end{equation}

For $\gamma < 3$ a strict cutoff $k_{\rm cut} = \sqrt{N}$ has been
introduced in the networks discussed in the present paper, in order 
to avoid undesired correlations \cite{ca05b}. Thus,
the equations given in this appendix do not apply to the actual 
networks discussed in Sec.~IV.C.


\begin{thebibliography}{57}
\expandafter\ifx\csname natexlab\endcsname\relax\def\natexlab#1{#1}\fi
\expandafter\ifx\csname bibnamefont\endcsname\relax
  \def\bibnamefont#1{#1}\fi
\expandafter\ifx\csname bibfnamefont\endcsname\relax
  \def\bibfnamefont#1{#1}\fi
\expandafter\ifx\csname citenamefont\endcsname\relax
  \def\citenamefont#1{#1}\fi
\expandafter\ifx\csname url\endcsname\relax
  \def\url#1{\texttt{#1}}\fi
\expandafter\ifx\csname urlprefix\endcsname\relax\def\urlprefix{URL }\fi
\providecommand{\bibinfo}[2]{#2}
\providecommand{\eprint}[2][]{\url{#2}}

\bibitem[{\citenamefont{Strogatz}(2001)}]{st01}
\bibinfo{author}{\bibfnamefont{S.~H.} \bibnamefont{Strogatz}},
  \bibinfo{journal}{Nature} \textbf{\bibinfo{volume}{410}},
  \bibinfo{pages}{268} (\bibinfo{year}{2001}).

\bibitem[{\citenamefont{Albert and Barab\'asi}(2002)}]{al02}
\bibinfo{author}{\bibfnamefont{R.}~\bibnamefont{Albert}} \bibnamefont{and}
  \bibinfo{author}{\bibfnamefont{A.~L.} \bibnamefont{Barab\'asi}},
  \bibinfo{journal}{Rev. Mod. Phys.} \textbf{\bibinfo{volume}{74}},
  \bibinfo{pages}{47} (\bibinfo{year}{2002}).

\bibitem[{\citenamefont{Dorogovtsev and Mendes}(2003)}]{do03}
\bibinfo{author}{\bibfnamefont{S.~N.} \bibnamefont{Dorogovtsev}}
  \bibnamefont{and} \bibinfo{author}{\bibfnamefont{J.~F.~F.}
  \bibnamefont{Mendes}}, \emph{\bibinfo{title}{Evolution of Networks: From
  Biological Nets to the Internet and WWW}} (\bibinfo{publisher}{Oxford
  University}, \bibinfo{address}{Oxford}, \bibinfo{year}{2003}).

\bibitem[{\citenamefont{Newman}(2010)}]{ne10}
\bibinfo{author}{\bibfnamefont{M.~E.~J.} \bibnamefont{Newman}},
  \emph{\bibinfo{title}{Networks. An Introduction}} (\bibinfo{publisher}{Oxford
  University Press}, \bibinfo{address}{New York}, \bibinfo{year}{2010}).

\bibitem[{\citenamefont{Cohen and Havlin}(2010)}]{co10}
\bibinfo{author}{\bibfnamefont{R.}~\bibnamefont{Cohen}} \bibnamefont{and}
  \bibinfo{author}{\bibfnamefont{S.}~\bibnamefont{Havlin}},
  \emph{\bibinfo{title}{Complex networks. Structure, robustness and function}}
  (\bibinfo{publisher}{Cambridge University Press},
  \bibinfo{address}{Cambridge}, \bibinfo{year}{2010}).

\bibitem[{\citenamefont{Watts and Strogatz}(1998)}]{wa98}
\bibinfo{author}{\bibfnamefont{D.~J.} \bibnamefont{Watts}} \bibnamefont{and}
  \bibinfo{author}{\bibfnamefont{S.~H.} \bibnamefont{Strogatz}},
  \bibinfo{journal}{Nature} \textbf{\bibinfo{volume}{393}},
  \bibinfo{pages}{440} (\bibinfo{year}{1998}).

\bibitem[{\citenamefont{Barab\'asi and Albert}(1999)}]{ba99b}
\bibinfo{author}{\bibfnamefont{A.~L.} \bibnamefont{Barab\'asi}}
  \bibnamefont{and} \bibinfo{author}{\bibfnamefont{R.}~\bibnamefont{Albert}},
  \bibinfo{journal}{Science} \textbf{\bibinfo{volume}{286}},
  \bibinfo{pages}{509} (\bibinfo{year}{1999}).

\bibitem[{\citenamefont{Herrero}(2002{\natexlab{a}})}]{he02b}
\bibinfo{author}{\bibfnamefont{C.~P.} \bibnamefont{Herrero}},
  \bibinfo{journal}{Phys. Rev. E} \textbf{\bibinfo{volume}{66}},
  \bibinfo{pages}{046126} (\bibinfo{year}{2002}{\natexlab{a}}).

\bibitem[{\citenamefont{Pandit and Amritkar}(2001)}]{pa01}
\bibinfo{author}{\bibfnamefont{S.~A.} \bibnamefont{Pandit}} \bibnamefont{and}
  \bibinfo{author}{\bibfnamefont{R.~E.} \bibnamefont{Amritkar}},
  \bibinfo{journal}{Phys. Rev. E} \textbf{\bibinfo{volume}{63}},
  \bibinfo{pages}{041104} (\bibinfo{year}{2001}).

\bibitem[{\citenamefont{Lahtinen et~al.}(2001)\citenamefont{Lahtinen,
  Kert\'esz, and Kaski}}]{la01b}
\bibinfo{author}{\bibfnamefont{J.}~\bibnamefont{Lahtinen}},
  \bibinfo{author}{\bibfnamefont{J.}~\bibnamefont{Kert\'esz}},
  \bibnamefont{and} \bibinfo{author}{\bibfnamefont{K.}~\bibnamefont{Kaski}},
  \bibinfo{journal}{Phys. Rev. E} \textbf{\bibinfo{volume}{64}},
  \bibinfo{pages}{057105} (\bibinfo{year}{2001}).

\bibitem[{\citenamefont{Candia}(2007)}]{ca07b}
\bibinfo{author}{\bibfnamefont{J.}~\bibnamefont{Candia}},
  \bibinfo{journal}{Phys. Rev. E} \textbf{\bibinfo{volume}{75}},
  \bibinfo{pages}{026110} (\bibinfo{year}{2007}).

\bibitem[{\citenamefont{Moore and Newman}(2000)}]{mo00}
\bibinfo{author}{\bibfnamefont{C.}~\bibnamefont{Moore}} \bibnamefont{and}
  \bibinfo{author}{\bibfnamefont{M.~E.~J.} \bibnamefont{Newman}},
  \bibinfo{journal}{Phys. Rev. E} \textbf{\bibinfo{volume}{61}},
  \bibinfo{pages}{5678} (\bibinfo{year}{2000}).

\bibitem[{\citenamefont{Kuperman and Abramson}(2001)}]{ku01}
\bibinfo{author}{\bibfnamefont{M.}~\bibnamefont{Kuperman}} \bibnamefont{and}
  \bibinfo{author}{\bibfnamefont{G.}~\bibnamefont{Abramson}},
  \bibinfo{journal}{Phys. Rev. Lett.} \textbf{\bibinfo{volume}{86}},
  \bibinfo{pages}{2909} (\bibinfo{year}{2001}).

\bibitem[{\citenamefont{Newman and Watts}(1999)}]{ne99}
\bibinfo{author}{\bibfnamefont{M.~E.~J.} \bibnamefont{Newman}}
  \bibnamefont{and} \bibinfo{author}{\bibfnamefont{D.~J.} \bibnamefont{Watts}},
  \bibinfo{journal}{Phys. Rev. E} \textbf{\bibinfo{volume}{60}},
  \bibinfo{pages}{7332} (\bibinfo{year}{1999}).

\bibitem[{\citenamefont{Barrat and Weigt}(2000)}]{ba00}
\bibinfo{author}{\bibfnamefont{A.}~\bibnamefont{Barrat}} \bibnamefont{and}
  \bibinfo{author}{\bibfnamefont{M.}~\bibnamefont{Weigt}},
  \bibinfo{journal}{Eur. Phys. J. B} \textbf{\bibinfo{volume}{13}},
  \bibinfo{pages}{547} (\bibinfo{year}{2000}).

\bibitem[{\citenamefont{Leone et~al.}(2002)\citenamefont{Leone, V\'azquez,
  Vespignani, and Zecchina}}]{le02}
\bibinfo{author}{\bibfnamefont{M.}~\bibnamefont{Leone}},
  \bibinfo{author}{\bibfnamefont{A.}~\bibnamefont{V\'azquez}},
  \bibinfo{author}{\bibfnamefont{A.}~\bibnamefont{Vespignani}},
  \bibnamefont{and} \bibinfo{author}{\bibfnamefont{R.}~\bibnamefont{Zecchina}},
  \bibinfo{journal}{Eur. Phys. J. B} \textbf{\bibinfo{volume}{28}},
  \bibinfo{pages}{191} (\bibinfo{year}{2002}).

\bibitem[{\citenamefont{Herrero}(2002{\natexlab{b}})}]{he02}
\bibinfo{author}{\bibfnamefont{C.~P.} \bibnamefont{Herrero}},
  \bibinfo{journal}{Phys. Rev. E} \textbf{\bibinfo{volume}{65}},
  \bibinfo{pages}{066110} (\bibinfo{year}{2002}{\natexlab{b}}).

\bibitem[{\citenamefont{Castellano et~al.}(2005)\citenamefont{Castellano,
  Loreto, Barrat, Cecconi, and Parisi}}]{ca05}
\bibinfo{author}{\bibfnamefont{C.}~\bibnamefont{Castellano}},
  \bibinfo{author}{\bibfnamefont{V.}~\bibnamefont{Loreto}},
  \bibinfo{author}{\bibfnamefont{A.}~\bibnamefont{Barrat}},
  \bibinfo{author}{\bibfnamefont{F.}~\bibnamefont{Cecconi}}, \bibnamefont{and}
  \bibinfo{author}{\bibfnamefont{D.}~\bibnamefont{Parisi}},
  \bibinfo{journal}{Phys. Rev. E} \textbf{\bibinfo{volume}{71}},
  \bibinfo{pages}{066107} (\bibinfo{year}{2005}).

\bibitem[{\citenamefont{Dorogovtsev et~al.}(2008)\citenamefont{Dorogovtsev,
  Goltsev, and Mendes}}]{do08}
\bibinfo{author}{\bibfnamefont{S.~N.} \bibnamefont{Dorogovtsev}},
  \bibinfo{author}{\bibfnamefont{A.~V.} \bibnamefont{Goltsev}},
  \bibnamefont{and} \bibinfo{author}{\bibfnamefont{J.~F.~F.}
  \bibnamefont{Mendes}}, \bibinfo{journal}{Rev. Mod. Phys.}
  \textbf{\bibinfo{volume}{80}}, \bibinfo{pages}{1275} (\bibinfo{year}{2008}).

\bibitem[{\citenamefont{Herrero}(2009{\natexlab{a}})}]{he09b}
\bibinfo{author}{\bibfnamefont{C.~P.} \bibnamefont{Herrero}},
  \bibinfo{journal}{J. Phys. A: Math. Gen.} \textbf{\bibinfo{volume}{42}},
  \bibinfo{pages}{415102} (\bibinfo{year}{2009}{\natexlab{a}}).

\bibitem[{\citenamefont{Ostilli et~al.}(2011)\citenamefont{Ostilli, Ferreira,
  and Mendes}}]{os11}
\bibinfo{author}{\bibfnamefont{M.}~\bibnamefont{Ostilli}},
  \bibinfo{author}{\bibfnamefont{A.~L.} \bibnamefont{Ferreira}},
  \bibnamefont{and} \bibinfo{author}{\bibfnamefont{J.~F.~F.}
  \bibnamefont{Mendes}}, \bibinfo{journal}{Phys. Rev. E}
  \textbf{\bibinfo{volume}{83}}, \bibinfo{pages}{061149}
  (\bibinfo{year}{2011}).

\bibitem[{\citenamefont{Dorogovtsev and Mendes}(2002)}]{do02}
\bibinfo{author}{\bibfnamefont{S.~N.} \bibnamefont{Dorogovtsev}}
  \bibnamefont{and} \bibinfo{author}{\bibfnamefont{J.~F.~F.}
  \bibnamefont{Mendes}}, \bibinfo{journal}{Adv. Phys.}
  \textbf{\bibinfo{volume}{51}}, \bibinfo{pages}{1079} (\bibinfo{year}{2002}).

\bibitem[{\citenamefont{Goh et~al.}(2002)\citenamefont{Goh, Oh, Jeong, Kahng,
  and Kim}}]{go02}
\bibinfo{author}{\bibfnamefont{K.~I.} \bibnamefont{Goh}},
  \bibinfo{author}{\bibfnamefont{E.~S.} \bibnamefont{Oh}},
  \bibinfo{author}{\bibfnamefont{H.}~\bibnamefont{Jeong}},
  \bibinfo{author}{\bibfnamefont{B.}~\bibnamefont{Kahng}}, \bibnamefont{and}
  \bibinfo{author}{\bibfnamefont{D.}~\bibnamefont{Kim}},
  \bibinfo{journal}{Proc. Natl. Acad. Sci. USA} \textbf{\bibinfo{volume}{99}},
  \bibinfo{pages}{12583} (\bibinfo{year}{2002}).

\bibitem[{\citenamefont{Siganos et~al.}(2003)\citenamefont{Siganos, Faloutsos,
  Faloutsos, and Faloutsos}}]{si03}
\bibinfo{author}{\bibfnamefont{G.}~\bibnamefont{Siganos}},
  \bibinfo{author}{\bibfnamefont{M.}~\bibnamefont{Faloutsos}},
  \bibinfo{author}{\bibfnamefont{P.}~\bibnamefont{Faloutsos}},
  \bibnamefont{and}
  \bibinfo{author}{\bibfnamefont{C.}~\bibnamefont{Faloutsos}},
  \bibinfo{journal}{IEEE ACM Trans. Netw.} \textbf{\bibinfo{volume}{11}},
  \bibinfo{pages}{514} (\bibinfo{year}{2003}).

\bibitem[{\citenamefont{Albert et~al.}(1999)\citenamefont{Albert, Jeong, and
  Barab\'asi}}]{al99}
\bibinfo{author}{\bibfnamefont{R.}~\bibnamefont{Albert}},
  \bibinfo{author}{\bibfnamefont{H.}~\bibnamefont{Jeong}}, \bibnamefont{and}
  \bibinfo{author}{\bibfnamefont{A.~L.} \bibnamefont{Barab\'asi}},
  \bibinfo{journal}{Nature} \textbf{\bibinfo{volume}{401}},
  \bibinfo{pages}{130} (\bibinfo{year}{1999}).

\bibitem[{\citenamefont{Jeong et~al.}(2001)\citenamefont{Jeong, Mason,
  Barab\'asi, and Oltvai}}]{je01}
\bibinfo{author}{\bibfnamefont{H.}~\bibnamefont{Jeong}},
  \bibinfo{author}{\bibfnamefont{S.~P.} \bibnamefont{Mason}},
  \bibinfo{author}{\bibfnamefont{A.~L.} \bibnamefont{Barab\'asi}},
  \bibnamefont{and} \bibinfo{author}{\bibfnamefont{Z.~N.}
  \bibnamefont{Oltvai}}, \bibinfo{journal}{Nature}
  \textbf{\bibinfo{volume}{411}}, \bibinfo{pages}{41} (\bibinfo{year}{2001}).

\bibitem[{\citenamefont{Newman}(2001)}]{ne01}
\bibinfo{author}{\bibfnamefont{M.~E.~J.} \bibnamefont{Newman}},
  \bibinfo{journal}{Proc. Natl. Acad. Sci. USA} \textbf{\bibinfo{volume}{98}},
  \bibinfo{pages}{404} (\bibinfo{year}{2001}).

\bibitem[{\citenamefont{Krapivsky et~al.}(2000)\citenamefont{Krapivsky, Redner,
  and Leyvraz}}]{kr00}
\bibinfo{author}{\bibfnamefont{P.~L.} \bibnamefont{Krapivsky}},
  \bibinfo{author}{\bibfnamefont{S.}~\bibnamefont{Redner}}, \bibnamefont{and}
  \bibinfo{author}{\bibfnamefont{F.}~\bibnamefont{Leyvraz}},
  \bibinfo{journal}{Phys. Rev. Lett.} \textbf{\bibinfo{volume}{85}},
  \bibinfo{pages}{4629} (\bibinfo{year}{2000}).

\bibitem[{\citenamefont{H\'ebert-Dufresne
  et~al.}(2011)\citenamefont{H\'ebert-Dufresne, Allard, Marceau, No\"el, and
  Dub\'e}}]{he11b}
\bibinfo{author}{\bibfnamefont{L.}~\bibnamefont{H\'ebert-Dufresne}},
  \bibinfo{author}{\bibfnamefont{A.}~\bibnamefont{Allard}},
  \bibinfo{author}{\bibfnamefont{V.}~\bibnamefont{Marceau}},
  \bibinfo{author}{\bibfnamefont{P.-A.} \bibnamefont{No\"el}},
  \bibnamefont{and} \bibinfo{author}{\bibfnamefont{L.~J.}
  \bibnamefont{Dub\'e}}, \bibinfo{journal}{Phys. Rev. Lett.}
  \textbf{\bibinfo{volume}{107}}, \bibinfo{pages}{158702}
  (\bibinfo{year}{2011}).

\bibitem[{\citenamefont{Bogacz et~al.}(2006)\citenamefont{Bogacz, Burda, and
  Waclaw}}]{bo06}
\bibinfo{author}{\bibfnamefont{L.}~\bibnamefont{Bogacz}},
  \bibinfo{author}{\bibfnamefont{Z.}~\bibnamefont{Burda}}, \bibnamefont{and}
  \bibinfo{author}{\bibfnamefont{B.}~\bibnamefont{Waclaw}},
  \bibinfo{journal}{Physica A} \textbf{\bibinfo{volume}{366}},
  \bibinfo{pages}{587} (\bibinfo{year}{2006}).

\bibitem[{\citenamefont{Holme and Kim}(2002)}]{ho02}
\bibinfo{author}{\bibfnamefont{P.}~\bibnamefont{Holme}} \bibnamefont{and}
  \bibinfo{author}{\bibfnamefont{B.~J.} \bibnamefont{Kim}},
  \bibinfo{journal}{Phys. Rev. E} \textbf{\bibinfo{volume}{65}},
  \bibinfo{pages}{026107} (\bibinfo{year}{2002}).

\bibitem[{\citenamefont{Klemm and Eguiluz}(2002)}]{kl02}
\bibinfo{author}{\bibfnamefont{K.}~\bibnamefont{Klemm}} \bibnamefont{and}
  \bibinfo{author}{\bibfnamefont{V.~M.} \bibnamefont{Eguiluz}},
  \bibinfo{journal}{Phys. Rev. E} \textbf{\bibinfo{volume}{65}},
  \bibinfo{pages}{036123} (\bibinfo{year}{2002}).

\bibitem[{\citenamefont{Serrano and Bogu\~n\'a}(2005)}]{se05}
\bibinfo{author}{\bibfnamefont{M.~A.} \bibnamefont{Serrano}} \bibnamefont{and}
  \bibinfo{author}{\bibfnamefont{M.}~\bibnamefont{Bogu\~n\'a}},
  \bibinfo{journal}{Phys. Rev. E} \textbf{\bibinfo{volume}{72}},
  \bibinfo{pages}{036133} (\bibinfo{year}{2005}).

\bibitem[{\citenamefont{Newman}(2009)}]{ne09}
\bibinfo{author}{\bibfnamefont{M.~E.~J.} \bibnamefont{Newman}},
  \bibinfo{journal}{Phys. Rev. Lett.} \textbf{\bibinfo{volume}{103}},
  \bibinfo{pages}{058701} (\bibinfo{year}{2009}).

\bibitem[{\citenamefont{Miller}(2009)}]{mi09}
\bibinfo{author}{\bibfnamefont{J.~C.} \bibnamefont{Miller}},
  \bibinfo{journal}{Phys. Rev. E} \textbf{\bibinfo{volume}{80}},
  \bibinfo{pages}{020901} (\bibinfo{year}{2009}).

\bibitem[{\citenamefont{Dorogovtsev et~al.}(2002)\citenamefont{Dorogovtsev,
  Goltsev, and Mendes}}]{do02b}
\bibinfo{author}{\bibfnamefont{S.~N.} \bibnamefont{Dorogovtsev}},
  \bibinfo{author}{\bibfnamefont{A.~V.} \bibnamefont{Goltsev}},
  \bibnamefont{and} \bibinfo{author}{\bibfnamefont{J.~F.~F.}
  \bibnamefont{Mendes}}, \bibinfo{journal}{Phys. Rev. E}
  \textbf{\bibinfo{volume}{66}}, \bibinfo{pages}{016104}
  (\bibinfo{year}{2002}).

\bibitem[{\citenamefont{Igl\'oi and Turban}(2002)}]{ig02}
\bibinfo{author}{\bibfnamefont{F.}~\bibnamefont{Igl\'oi}} \bibnamefont{and}
  \bibinfo{author}{\bibfnamefont{L.}~\bibnamefont{Turban}},
  \bibinfo{journal}{Phys. Rev. E} \textbf{\bibinfo{volume}{66}},
  \bibinfo{pages}{036140} (\bibinfo{year}{2002}).

\bibitem[{\citenamefont{Herrero}(2004)}]{he04}
\bibinfo{author}{\bibfnamefont{C.~P.} \bibnamefont{Herrero}},
  \bibinfo{journal}{Phys. Rev. E} \textbf{\bibinfo{volume}{69}},
  \bibinfo{pages}{067109} (\bibinfo{year}{2004}).

\bibitem[{\citenamefont{Dommers et~al.}(2010)\citenamefont{Dommers, Giardina,
  and van~der Hofstad}}]{do10}
\bibinfo{author}{\bibfnamefont{S.}~\bibnamefont{Dommers}},
  \bibinfo{author}{\bibfnamefont{C.}~\bibnamefont{Giardina}}, \bibnamefont{and}
  \bibinfo{author}{\bibfnamefont{R.}~\bibnamefont{van~der Hofstad}},
  \bibinfo{journal}{J. Stat. Phys.} \textbf{\bibinfo{volume}{141}},
  \bibinfo{pages}{638} (\bibinfo{year}{2010}).

\bibitem[{\citenamefont{Menche et~al.}(2011)\citenamefont{Menche, Valleriani,
  and Lipowsky}}]{me11b}
\bibinfo{author}{\bibfnamefont{J.}~\bibnamefont{Menche}},
  \bibinfo{author}{\bibfnamefont{A.}~\bibnamefont{Valleriani}},
  \bibnamefont{and} \bibinfo{author}{\bibfnamefont{R.}~\bibnamefont{Lipowsky}},
  \bibinfo{journal}{Phys. Rev. E} \textbf{\bibinfo{volume}{83}},
  \bibinfo{pages}{061129} (\bibinfo{year}{2011}).

\bibitem[{\citenamefont{Bartolozzi et~al.}(2006)\citenamefont{Bartolozzi,
  Surungan, Leinweber, and Williams}}]{ba06}
\bibinfo{author}{\bibfnamefont{M.}~\bibnamefont{Bartolozzi}},
  \bibinfo{author}{\bibfnamefont{T.}~\bibnamefont{Surungan}},
  \bibinfo{author}{\bibfnamefont{D.~B.} \bibnamefont{Leinweber}},
  \bibnamefont{and} \bibinfo{author}{\bibfnamefont{A.~G.}
  \bibnamefont{Williams}}, \bibinfo{journal}{Phys. Rev. B}
  \textbf{\bibinfo{volume}{73}}, \bibinfo{pages}{224419}
  (\bibinfo{year}{2006}).

\bibitem[{\citenamefont{Herrero}(2009{\natexlab{b}})}]{he09}
\bibinfo{author}{\bibfnamefont{C.~P.} \bibnamefont{Herrero}},
  \bibinfo{journal}{Eur. Phys. J. B} \textbf{\bibinfo{volume}{70}},
  \bibinfo{pages}{435} (\bibinfo{year}{2009}{\natexlab{b}}).

\bibitem[{\citenamefont{Gleeson}(2009)}]{gl09}
\bibinfo{author}{\bibfnamefont{J.~P.} \bibnamefont{Gleeson}},
  \bibinfo{journal}{Phys. Rev. E} \textbf{\bibinfo{volume}{80}},
  \bibinfo{pages}{036107} (\bibinfo{year}{2009}).

\bibitem[{\citenamefont{Gleeson et~al.}(2010)\citenamefont{Gleeson, Melnik, and
  Hackett}}]{gl10}
\bibinfo{author}{\bibfnamefont{J.~P.} \bibnamefont{Gleeson}},
  \bibinfo{author}{\bibfnamefont{S.}~\bibnamefont{Melnik}}, \bibnamefont{and}
  \bibinfo{author}{\bibfnamefont{A.}~\bibnamefont{Hackett}},
  \bibinfo{journal}{Phys. Rev. E} \textbf{\bibinfo{volume}{81}},
  \bibinfo{pages}{066114} (\bibinfo{year}{2010}).

\bibitem[{\citenamefont{Karrer and Newman}(2010)}]{ka10b}
\bibinfo{author}{\bibfnamefont{B.}~\bibnamefont{Karrer}} \bibnamefont{and}
  \bibinfo{author}{\bibfnamefont{M.~E.~J.} \bibnamefont{Newman}},
  \bibinfo{journal}{Phys. Rev. E} \textbf{\bibinfo{volume}{82}},
  \bibinfo{pages}{066118} (\bibinfo{year}{2010}).

\bibitem[{\citenamefont{Allard et~al.}(2012)\citenamefont{Allard,
  Hebert-Dufresne, Noel, Marceau, and Dube}}]{al12}
\bibinfo{author}{\bibfnamefont{A.}~\bibnamefont{Allard}},
  \bibinfo{author}{\bibfnamefont{L.}~\bibnamefont{Hebert-Dufresne}},
  \bibinfo{author}{\bibfnamefont{P.-A.} \bibnamefont{Noel}},
  \bibinfo{author}{\bibfnamefont{V.}~\bibnamefont{Marceau}}, \bibnamefont{and}
  \bibinfo{author}{\bibfnamefont{L.~J.} \bibnamefont{Dube}},
  \bibinfo{journal}{J. Phys. A: Math. Theor.} \textbf{\bibinfo{volume}{45}},
  \bibinfo{pages}{405005} (\bibinfo{year}{2012}).

\bibitem[{\citenamefont{Wang et~al.}(2012)\citenamefont{Wang, Cao, Suzuki, and
  Aihara}}]{wa12}
\bibinfo{author}{\bibfnamefont{B.}~\bibnamefont{Wang}},
  \bibinfo{author}{\bibfnamefont{L.}~\bibnamefont{Cao}},
  \bibinfo{author}{\bibfnamefont{H.}~\bibnamefont{Suzuki}}, \bibnamefont{and}
  \bibinfo{author}{\bibfnamefont{K.}~\bibnamefont{Aihara}},
  \bibinfo{journal}{J. Theor. Biol.} \textbf{\bibinfo{volume}{304}},
  \bibinfo{pages}{121} (\bibinfo{year}{2012}).

\bibitem[{\citenamefont{Molina and Stone}(2012)}]{mo12}
\bibinfo{author}{\bibfnamefont{C.}~\bibnamefont{Molina}} \bibnamefont{and}
  \bibinfo{author}{\bibfnamefont{L.}~\bibnamefont{Stone}}, \bibinfo{journal}{J.
  Theor. Biol.} \textbf{\bibinfo{volume}{315}}, \bibinfo{pages}{110}
  (\bibinfo{year}{2012}).

\bibitem[{\citenamefont{Hebert-Dufresne
  et~al.}(2010)\citenamefont{Hebert-Dufresne, Noel, Marceau, Allard, and
  Dube}}]{he10}
\bibinfo{author}{\bibfnamefont{L.}~\bibnamefont{Hebert-Dufresne}},
  \bibinfo{author}{\bibfnamefont{P.-A.} \bibnamefont{Noel}},
  \bibinfo{author}{\bibfnamefont{V.}~\bibnamefont{Marceau}},
  \bibinfo{author}{\bibfnamefont{A.}~\bibnamefont{Allard}}, \bibnamefont{and}
  \bibinfo{author}{\bibfnamefont{L.~J.} \bibnamefont{Dube}},
  \bibinfo{journal}{Phys. Rev. E} \textbf{\bibinfo{volume}{82}},
  \bibinfo{pages}{036115} (\bibinfo{year}{2010}).

\bibitem[{\citenamefont{Melnik et~al.}(2011)\citenamefont{Melnik, Hackett,
  Porter, Mucha, and Gleeson}}]{me11}
\bibinfo{author}{\bibfnamefont{S.}~\bibnamefont{Melnik}},
  \bibinfo{author}{\bibfnamefont{A.}~\bibnamefont{Hackett}},
  \bibinfo{author}{\bibfnamefont{M.~A.} \bibnamefont{Porter}},
  \bibinfo{author}{\bibfnamefont{P.~J.} \bibnamefont{Mucha}}, \bibnamefont{and}
  \bibinfo{author}{\bibfnamefont{J.~P.} \bibnamefont{Gleeson}},
  \bibinfo{journal}{Phys. Rev. E} \textbf{\bibinfo{volume}{83}},
  \bibinfo{pages}{036112} (\bibinfo{year}{2011}).

\bibitem[{\citenamefont{Yoon et~al.}(2011)\citenamefont{Yoon, Goltsev,
  Dorogovtsev, and Mendes}}]{yo11}
\bibinfo{author}{\bibfnamefont{S.}~\bibnamefont{Yoon}},
  \bibinfo{author}{\bibfnamefont{A.~V.} \bibnamefont{Goltsev}},
  \bibinfo{author}{\bibfnamefont{S.~N.} \bibnamefont{Dorogovtsev}},
  \bibnamefont{and} \bibinfo{author}{\bibfnamefont{J.~F.~F.}
  \bibnamefont{Mendes}}, \bibinfo{journal}{Phys. Rev. E}
  \textbf{\bibinfo{volume}{84}}, \bibinfo{pages}{041144}
  (\bibinfo{year}{2011}).

\bibitem[{\citenamefont{Newman}(2005)}]{ne05}
\bibinfo{author}{\bibfnamefont{M.~E.~J.} \bibnamefont{Newman}},
  \bibinfo{journal}{Contemp. Phys.} \textbf{\bibinfo{volume}{46}},
  \bibinfo{pages}{323} (\bibinfo{year}{2005}).

\bibitem[{\citenamefont{Catanzaro et~al.}(2005)\citenamefont{Catanzaro,
  Bogu\~n\'a, and Pastor-Satorras}}]{ca05b}
\bibinfo{author}{\bibfnamefont{M.}~\bibnamefont{Catanzaro}},
  \bibinfo{author}{\bibfnamefont{M.}~\bibnamefont{Bogu\~n\'a}},
  \bibnamefont{and}
  \bibinfo{author}{\bibfnamefont{R.}~\bibnamefont{Pastor-Satorras}},
  \bibinfo{journal}{Phys. Rev. E} \textbf{\bibinfo{volume}{71}},
  \bibinfo{pages}{027103} (\bibinfo{year}{2005}).

\bibitem[{\citenamefont{Binder and Heermann}(2010)}]{bi10}
\bibinfo{author}{\bibfnamefont{K.}~\bibnamefont{Binder}} \bibnamefont{and}
  \bibinfo{author}{\bibfnamefont{D.~W.} \bibnamefont{Heermann}},
  \emph{\bibinfo{title}{Monte Carlo Simulation in Statistical Physics}}
  (\bibinfo{publisher}{Springer}, \bibinfo{address}{Berlin},
  \bibinfo{year}{2010}), \bibinfo{edition}{5th} ed.

\bibitem[{\citenamefont{Herrero}(2008)}]{he08}
\bibinfo{author}{\bibfnamefont{C.~P.} \bibnamefont{Herrero}},
  \bibinfo{journal}{Phys. Rev. E} \textbf{\bibinfo{volume}{77}},
  \bibinfo{pages}{041102} (\bibinfo{year}{2008}).

\bibitem[{\citenamefont{Aleksiejuk et~al.}(2002)\citenamefont{Aleksiejuk,
  Holyst, and Stauffer}}]{al02b}
\bibinfo{author}{\bibfnamefont{A.}~\bibnamefont{Aleksiejuk}},
  \bibinfo{author}{\bibfnamefont{J.~A.} \bibnamefont{Holyst}},
  \bibnamefont{and} \bibinfo{author}{\bibfnamefont{D.}~\bibnamefont{Stauffer}},
  \bibinfo{journal}{Physica A} \textbf{\bibinfo{volume}{310}},
  \bibinfo{pages}{260} (\bibinfo{year}{2002}).

\bibitem[{\citenamefont{Aleksiejuk-Fronczak}(2002)}]{al02c}
\bibinfo{author}{\bibfnamefont{A.}~\bibnamefont{Aleksiejuk-Fronczak}},
  \bibinfo{journal}{Int. J. Mod. Phys. C} \textbf{\bibinfo{volume}{13}},
  \bibinfo{pages}{1415} (\bibinfo{year}{2002}).

\end{thebibliography}
\end{document}